\documentclass[aps,pre,amsmath,superscriptaddress,amssymb,twocolumn,floatfix]{revtex4}     
\usepackage{graphicx}
\usepackage{color}

\begin{document}
\title{``Swarm relaxation":  Equilibrating a large ensemble of computer simulations}

\author{Shahrazad M.A. Malek}
\affiliation{Department of Physics and Physical Oceanography, Memorial University of Newfoundland, \\St. John's, Newfoundland A1B 3X7, Canada}

\author{Richard K. Bowles}
\affiliation{Department of Chemistry, University of Saskatchewan, Saskatoon, SK, 57N 5C9, Canada}

\author{Ivan Saika-Voivod}
\affiliation{Department of Physics and Physical Oceanography, Memorial University of Newfoundland, \\St. John's, Newfoundland A1B 3X7, Canada}

\author{Francesco Sciortino}
\affiliation{Dipartimento di Fisica,
Universit\`a di Roma {\em La Sapienza},
Piazzale A. Moro 5, 00185 Roma, Italy}

\author{Peter H. Poole}
\affiliation{Department of Physics, St. Francis Xavier University, Antigonish, NS, B2G 2W5, Canada}

\begin{abstract}

It is common practice in molecular dynamics and Monte Carlo computer simulations to run multiple, separately-initialized simulations in order to improve the sampling of independent microstates.  Here we examine the utility of an extreme case of this strategy, in which we run
a large ensemble of $M$ independent simulations (a ``swarm"), each of which is relaxed to equilibrium.
We show that if $M$ is of order $10^3$, we can monitor the swarm's relaxation to equilibrium, and confirm its attainment, within $\sim 10\bar\tau$, where $\bar\tau$ is the equilibrium relaxation time.  As soon as a swarm of this size attains equilibrium, the ensemble of $M$ final microstates from each run 
is sufficient for
the evaluation of most equilibrium properties without further sampling.  
This approach dramatically reduces the wall-clock time required, compared to a single long simulation, by a factor of several hundred, 
at the cost of an increase in the total computational effort by a small factor.
It is also well-suited to modern computing systems having thousands of processors, and is a viable strategy for simulation studies that need to produce 
high-precision results in a minimum of wall-clock time.
We present results obtained by applying this approach to several test cases.
\end{abstract}
\date{\today\,  -- Eur. Phys. J. E, in press, 2017}
\maketitle

\begin{table*}
  \begin{tabular}{ | c | c | c | c | c | c | c | c | c |}
    \hline
    swarm & $T$ & $P$ & started & $t_{\rm run}$ & $\bar \tau$ & $t_{\rm run}/{\bar \tau}$ & $t_{\rm stop}$ &$t_{\rm stop}/\bar\tau$ \\ 
    run label & (K) & (MPa) & from & ($10^3$ MCS or ns) & ($10^3$ MCS or ns) & & ($10^3$ MCS or ns) & \\ \hline \hline
    A & 400 & 100 & random & $100 $ & $3.4$ & 29 & 41 & 12\\ 	
    B & 290 & 120 & A & $400 $ & $19$& 21 & 229 & 12 \\ 
    C & 250 & 190 & A & $4000$& $230$& 17 & 3270 & 14\\ 
    D & 100 & 190 & C & $800$& $\gg 10^3$& $\ll 1$ &$\gg 10^3$ & - \\ \hline
    E & 180 & - & SLR at 180 K & $24$& $1.8$& $13$ &$23$ & 13 \\ 
    F & 180 & - & SLR at 220 K & $24$& $1.9$& $13$ &$23$ & 12 \\ 
     \hline
  \end{tabular}
  \caption{Run parameters and time scales for each of our swarm relaxation test cases.
  Symbols and abbreviations are as defined in the text.  Time units are MCS for runs A, B, C and D, and are ns for E and F.}
  \label{table}
  \end{table*}

\section{Introduction}

When conducting a molecular dynamics or Monte Carlo computer simulation study of an equilibrium system, a key question is: ``How long should we run?"  First, equilibrium must be attained and verified, and then a sufficient number of independent microstates of the system must be sampled within equilibrium to allow for the accurate evaluation of equilibrium properties.  In a traditional approach, all of this is achieved in a single long run (SLR).  In this context, a run is ``long" if it is many times (usually 100 times or more) longer than the equilibrium relaxation time $\bar\tau$ of the slowest relaxing, unconstrained observable of the system.  When using a SLR, the evaluation of equilibrium properties relies on the ergodic hypothesis, i.e. that a sufficiently long time average of an observable is equal to the ensemble average taken over a set of independently generated microstates~\cite{frenkel}.

While perfectly sound in principle, a SLR can produce inaccurate results if $\bar\tau$ is underestimated.  This can occur in simulations of supercooled liquids and glassy systems exhibiting subtle and very slow structural relaxation~\cite{berthier}, or in complex systems (such as proteins) where metastable basins of the free energy landscape trap the system for time scales that are long compared to the time required to explore the metastable basin itself~\cite{karplus}.  In these cases, a SLR may appear to achieve equilibrium when in fact it has not.

As a consequence of these concerns, it is increasingly common to initiate multiple, independently initialized simulation runs to test for slow relaxation and trapping in metastable states~\cite{nilsson,karplus,cov1,cov2}.  This strategy also takes advantage of the multi-processor structure of virtually all modern computing systems, since independent simulations can run concurrently on separate processors.  Simulation studies of aging in glassy materials have long used this approach, in order to average over different realizations of the disorder in the initial configuration~\cite{kob,la2006relation}.  

When using multiple runs to study an equilibrium system, the final results are averaged both in time (within a single run) and over the ensemble of independent runs.  Here we study the extreme case of an ensemble of runs in which the number of runs $M$ is so large that no time averaging is required to obtain accurate results.  Herein, we refer to such a large ensemble of runs as a ``swarm". That is, we create a swarm of $M$ independent runs, bring each to equilibrium, and use only the last microstate of each run to evaluate the equilibirum properties, which are computed purely as ensemble averages.  

Our motivation to study this extreme case is to minimize the wall-clock time required to obtain the final results:  The shortest possible run that produces an equilibrium microstate is a run that just reaches equilibrium and then stops.  If a swarm of $M$ such runs is carried out concurrently, and if $M$ is large enough to produce an accurate ensemble average, then the wall-clock time to obtain results of a given precision will be substantially less than for a SLR.  While it is apparent that this strategy can produce accurate results if $M$ is large enough, and if the runs are long enough, it is not obvious that the reduction in the wall-clock time will be worth the increase in the total computational cost, compared to a SLR.  The efficiency of such a ``swarm relaxation" strategy, relative to a SLR, will depend on the ability to stop the swarm runs just as they relax to equilibrium.  However, we usually don't know the time scale to reach equilibrium in advance.  

In the following, we study several test cases of the swarm relaxation approach, using Monte Carlo and molecular dynamics simulations of water.  Simulations of water display a wealth of complex phenomena, carefully studied in many previous works, making this system an excellent choice for testing new computational strategies.
We test the swarm relaxation approach by examining the time dependence of average properties, and their variance, during the evolution of the swarm to equilibrium, and also examine the properties of the autocorrelation functions and relaxation times of these observables.  For several test cases, we show that when $M$ is large enough (of order $10^3$ or greater), the establishment of equilibrium can be detected from the time evolution of the average properties of the swarm on a time scale which is not much longer than the time scale separating independent equilibrium microstates in a single run.   We also show that such values of $M$ are sufficient to accurately evaluate equilibrium properties.  For our test cases, when all $M$ simulations in the swarm run concurrently, we show that a dramatic decrease of the wall-clock time is achieved (a factor of several hundred), in return for a much smaller increase in the total computational cost (a factor of not more than 3), relative to a SLR.  Thus a swarm relaxation strategy is a viable approach for exploiting large-scale multi-processor computing systems to substantially reduce the wall-clock time required to evaluate equilibrium properties.

\begin{figure*}
{\includegraphics[scale=0.35]{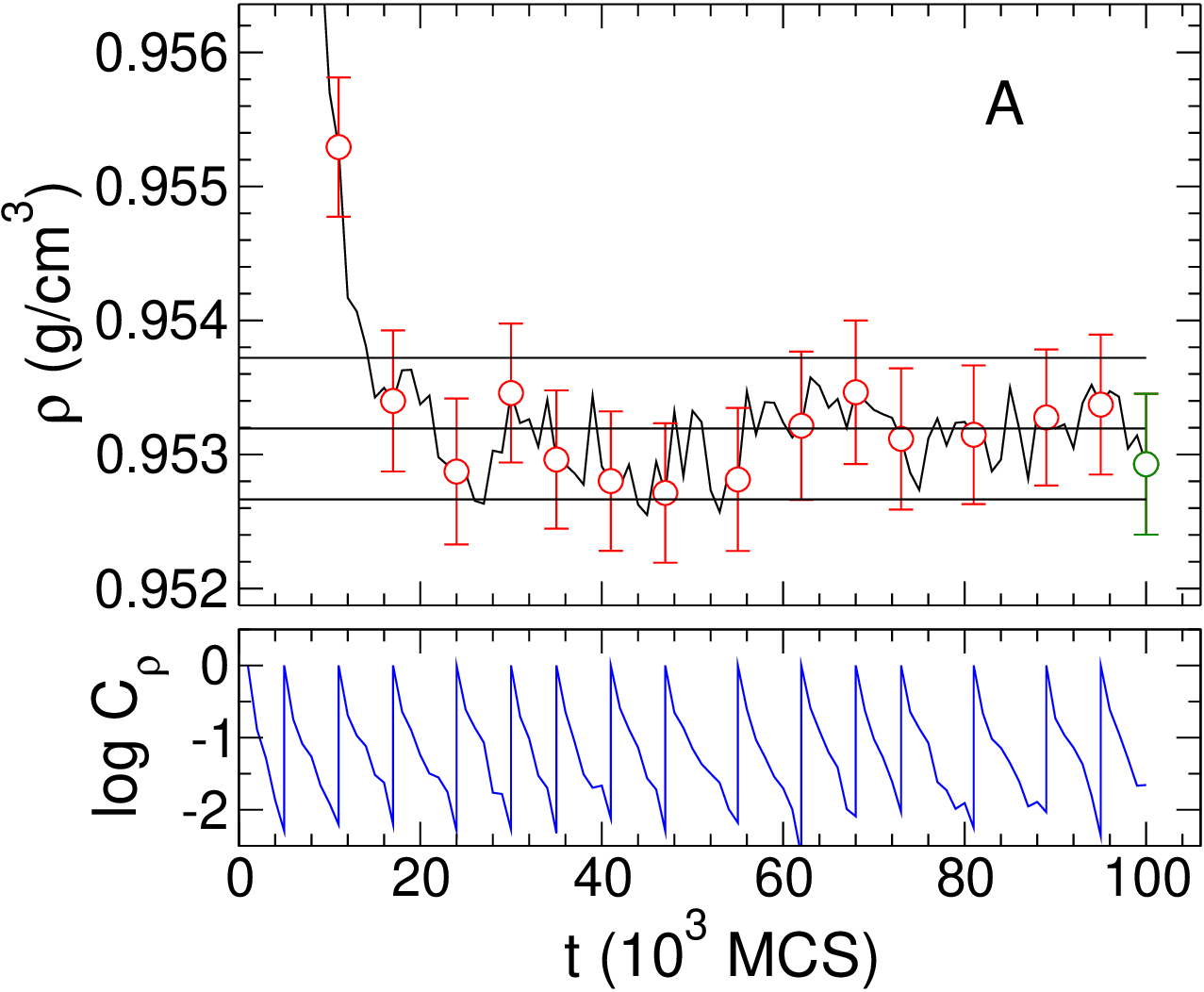}}\hskip 0.5 in
{\includegraphics[scale=0.35]{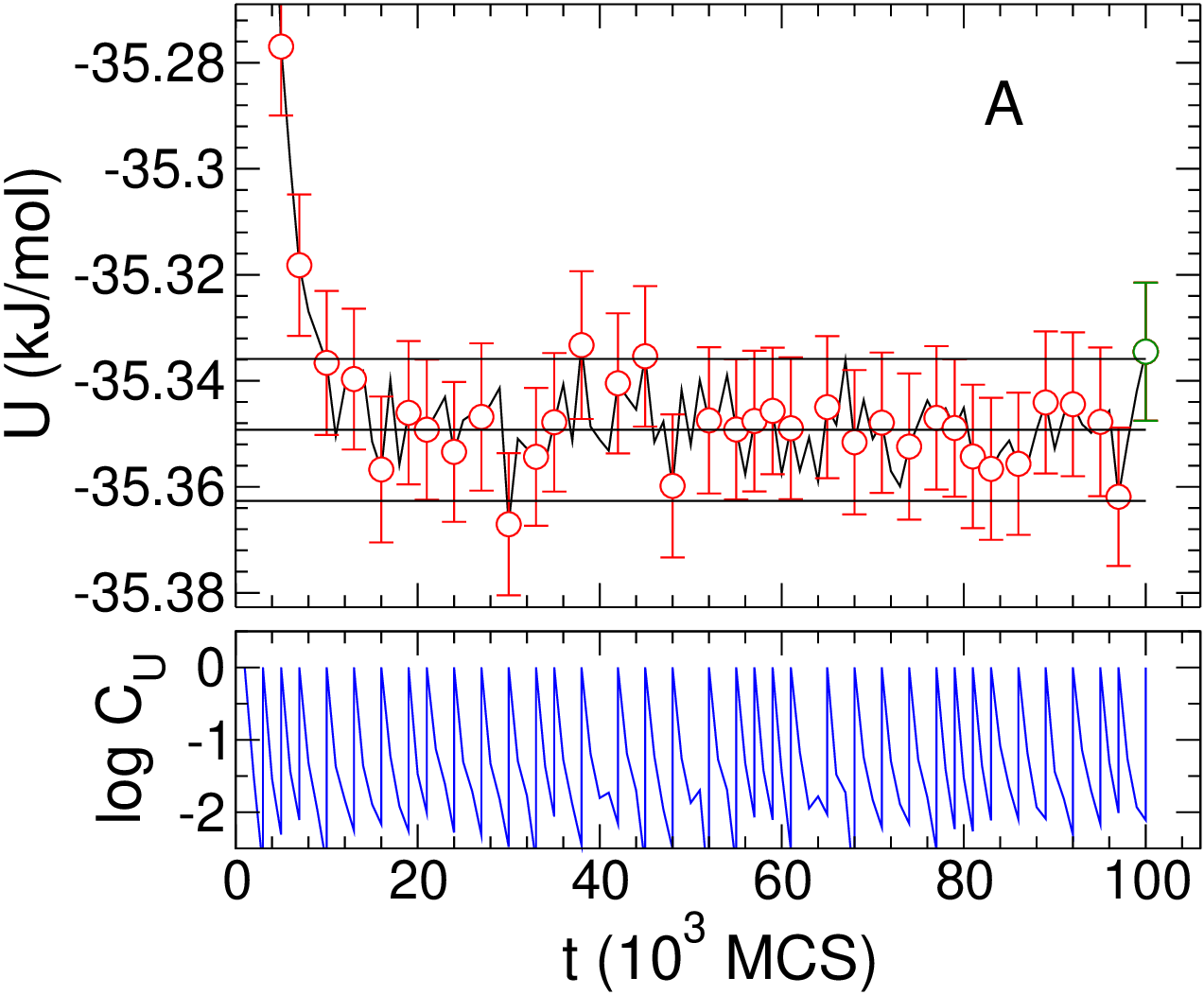}}\vskip 0.4 in
{\includegraphics[scale=0.35]{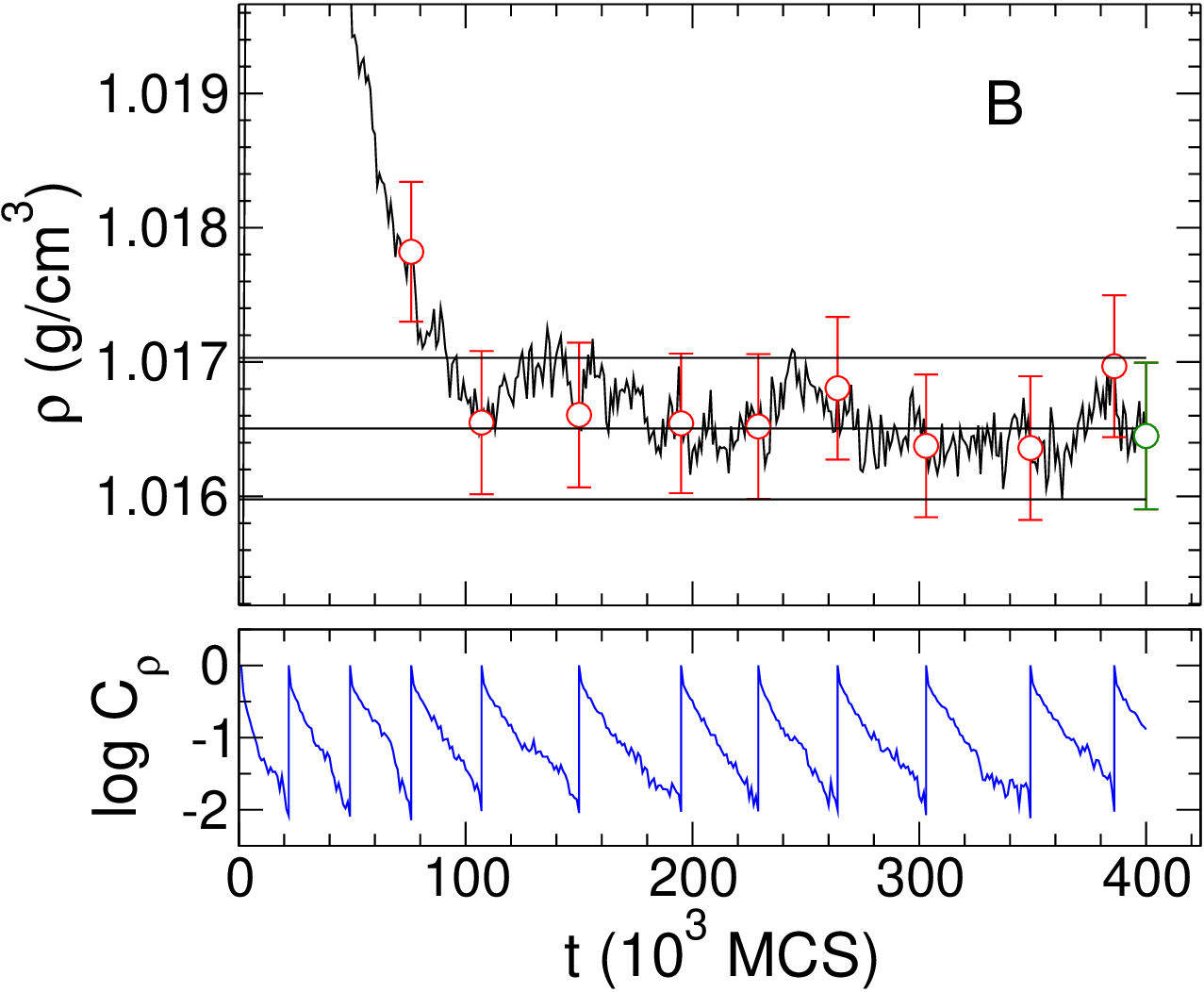}}\hskip 0.5 in
{\includegraphics[scale=0.35]{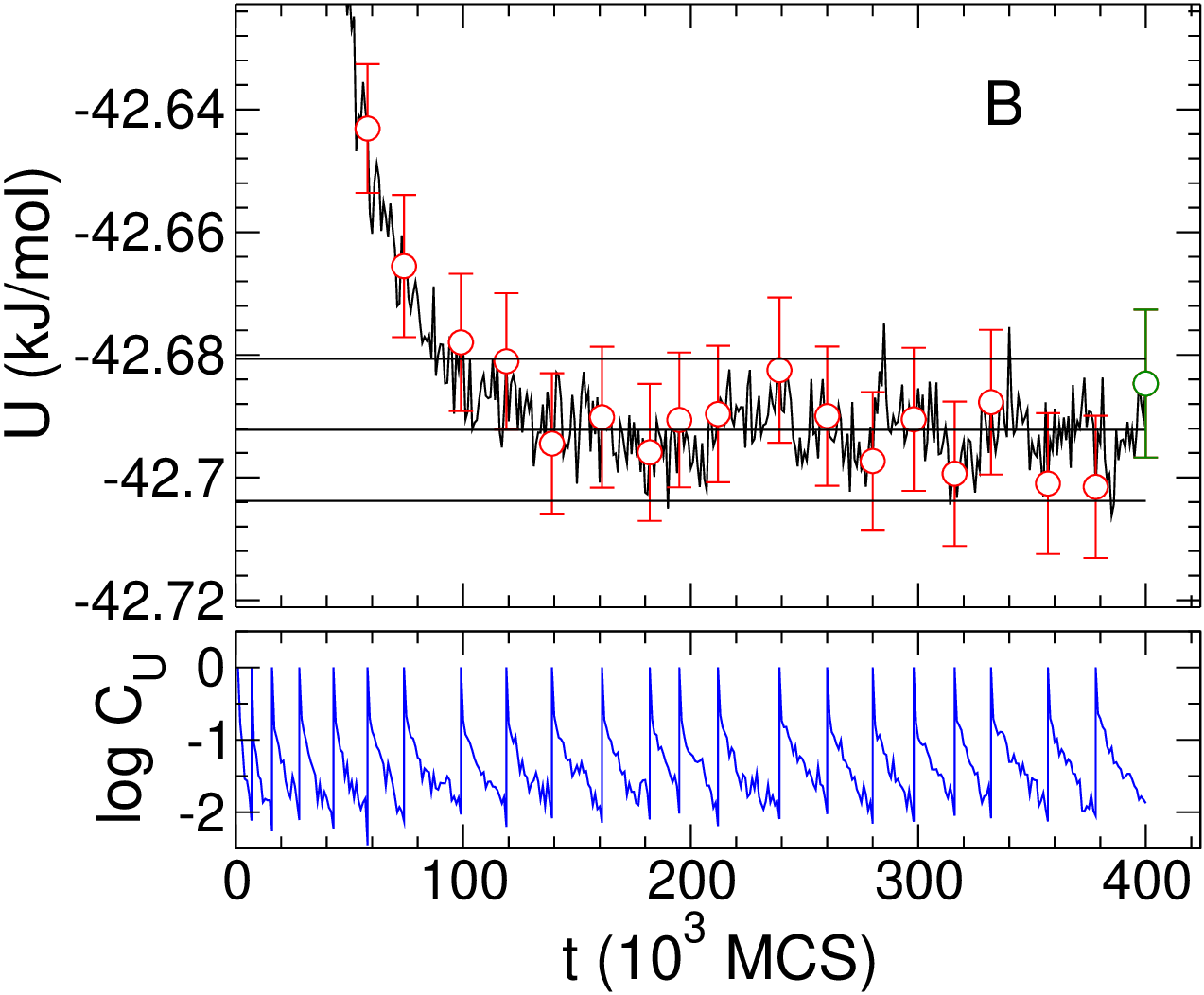}}\vskip 0.4 in
{\includegraphics[scale=0.35]{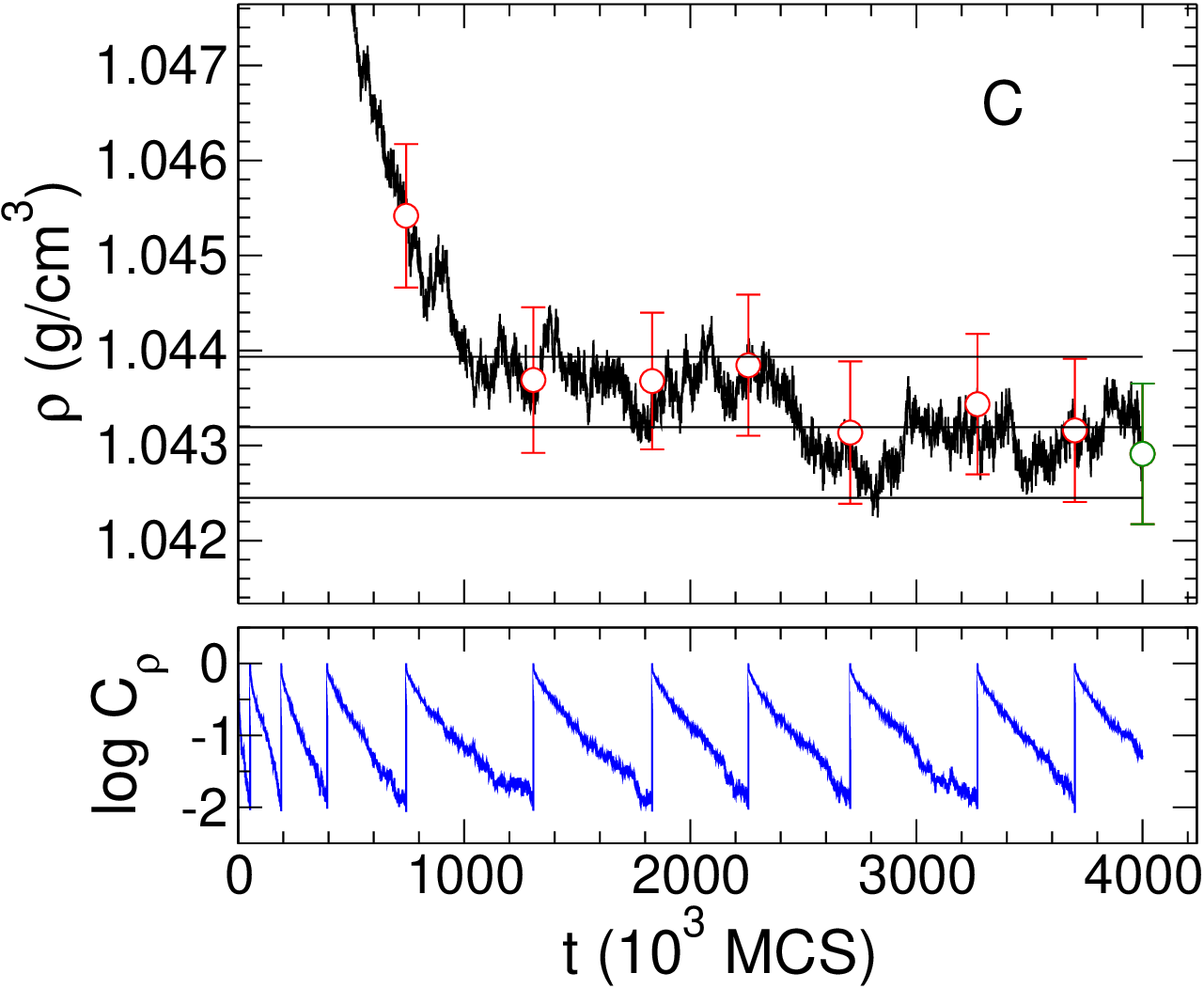}}\hskip 0.5 in
{\includegraphics[scale=0.35]{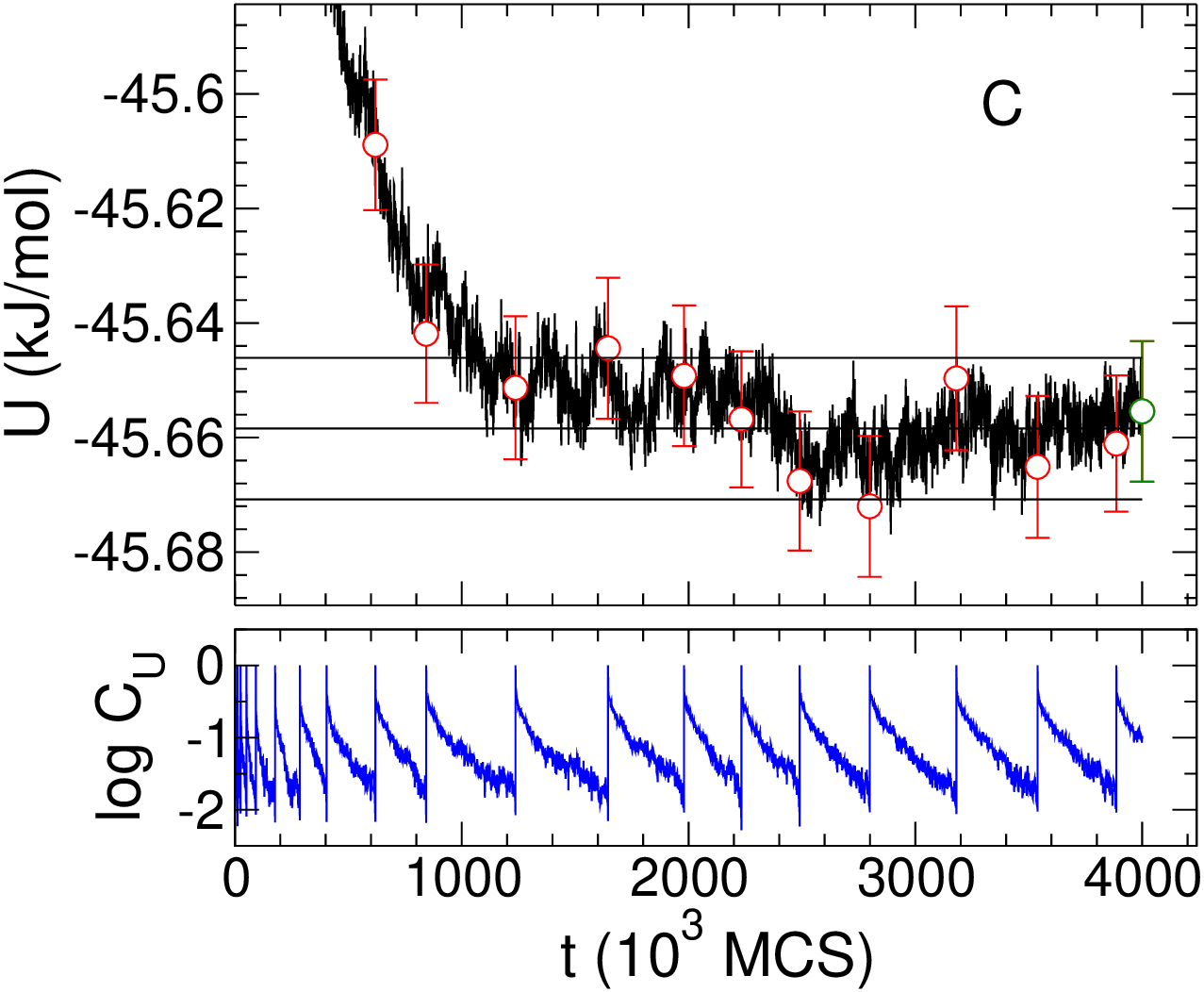}}
\caption{Black curves show the time dependence of $\langle \rho \rangle$ (left panels)
and $\langle U \rangle$ (right panels)
for ST2 runs A, B, and C (panels top to bottom).
The black horizontal lines identify $\bar \rho$ and $\bar \rho \pm 2 \bar s_\rho$ (left panels); 
and $\bar U$ and $\bar \rho \pm 2 \bar s_U$ (right panels).  
The bottom section of each panel shows $\log C_\rho$ (left panels) and 
$\log C_U$ (right panels)
over successive relaxation cycles, calculated as described in the text.  
The red circles in the left panels are values of $\langle \rho \rangle$ (with error $\pm 2 s_\rho$) at the beginning of each relaxation cycle, and the green circle is $\langle \rho \rangle$ at $t=t_{\rm run}$.
Similarly, the red circles in the right panels are values of $\langle U \rangle$ (with error $\pm 2 s_U$) at the beginning of each relaxation cycle, and the green circle is $\langle U \rangle$ at $t=t_{\rm run}$.
}
\label{st2}
\end{figure*}

\begin{figure}\bigskip\bigskip
\centerline{\includegraphics[scale=0.35]{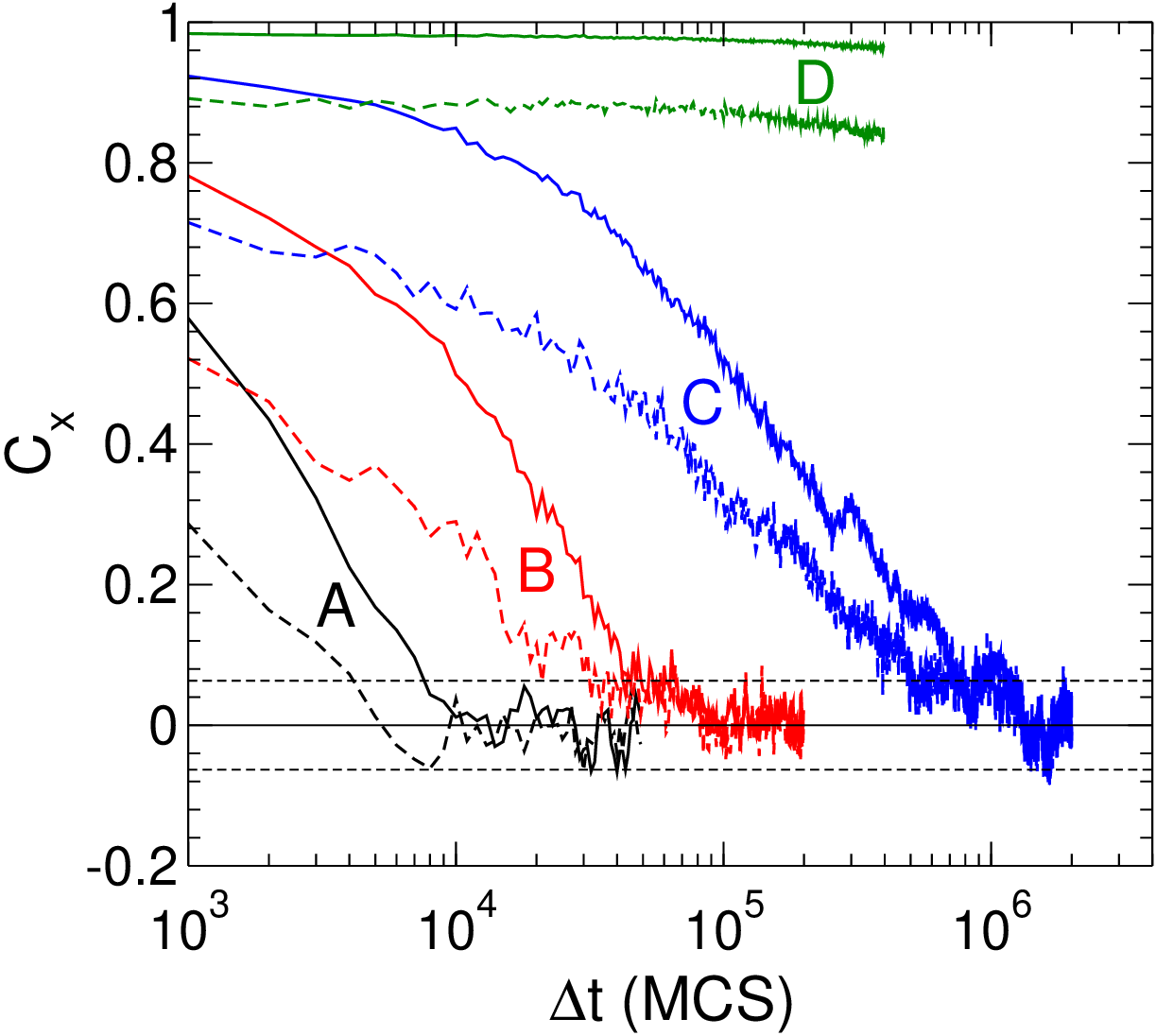}}
\caption{$C_\rho$ (solid lines) and $C_U$ (dashed lines) for ST2 runs A, B, C and D.
The horizontal dotted lines identify $C_x=\pm 2M^{-1/2}$.  Here $\Delta t=t-t_0$, with $t_0=t_{\rm run}/2$.  
}
\label{C}
\end{figure}

\begin{figure}\bigskip\bigskip
\centerline{\includegraphics[scale=0.35]{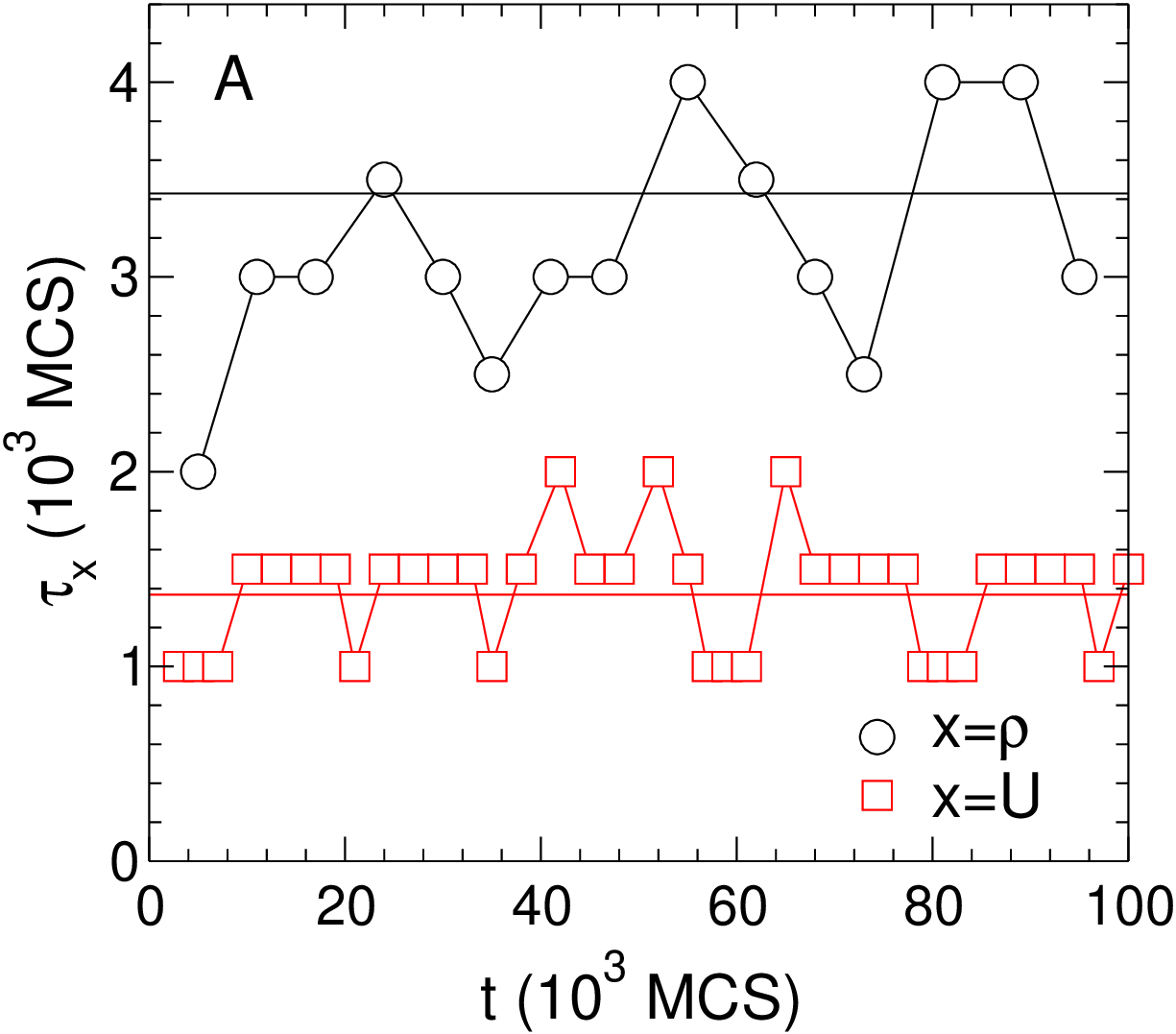}}\vskip 0.4 in
\centerline{\includegraphics[scale=0.35]{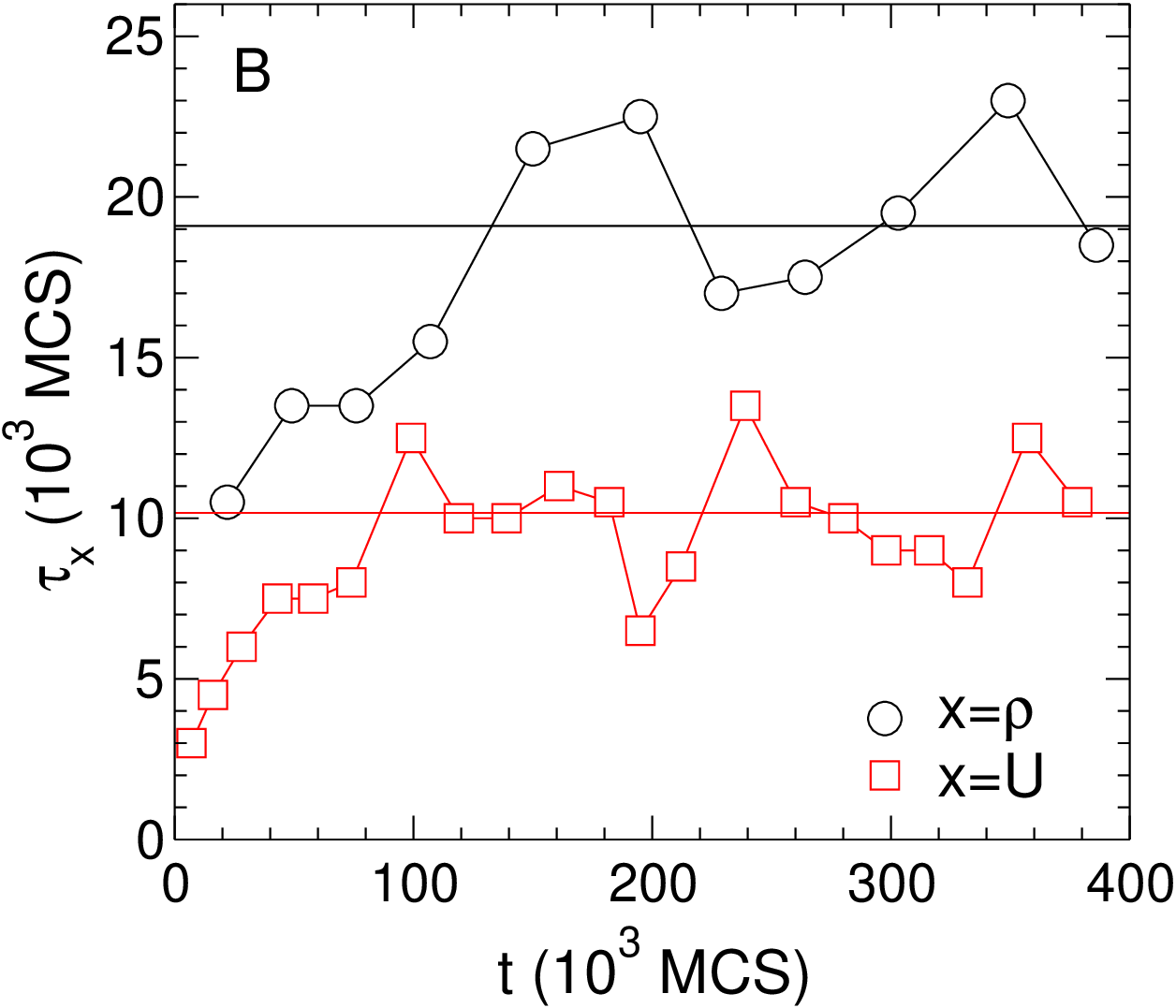}}\vskip 0.4 in
\centerline{\includegraphics[scale=0.35]{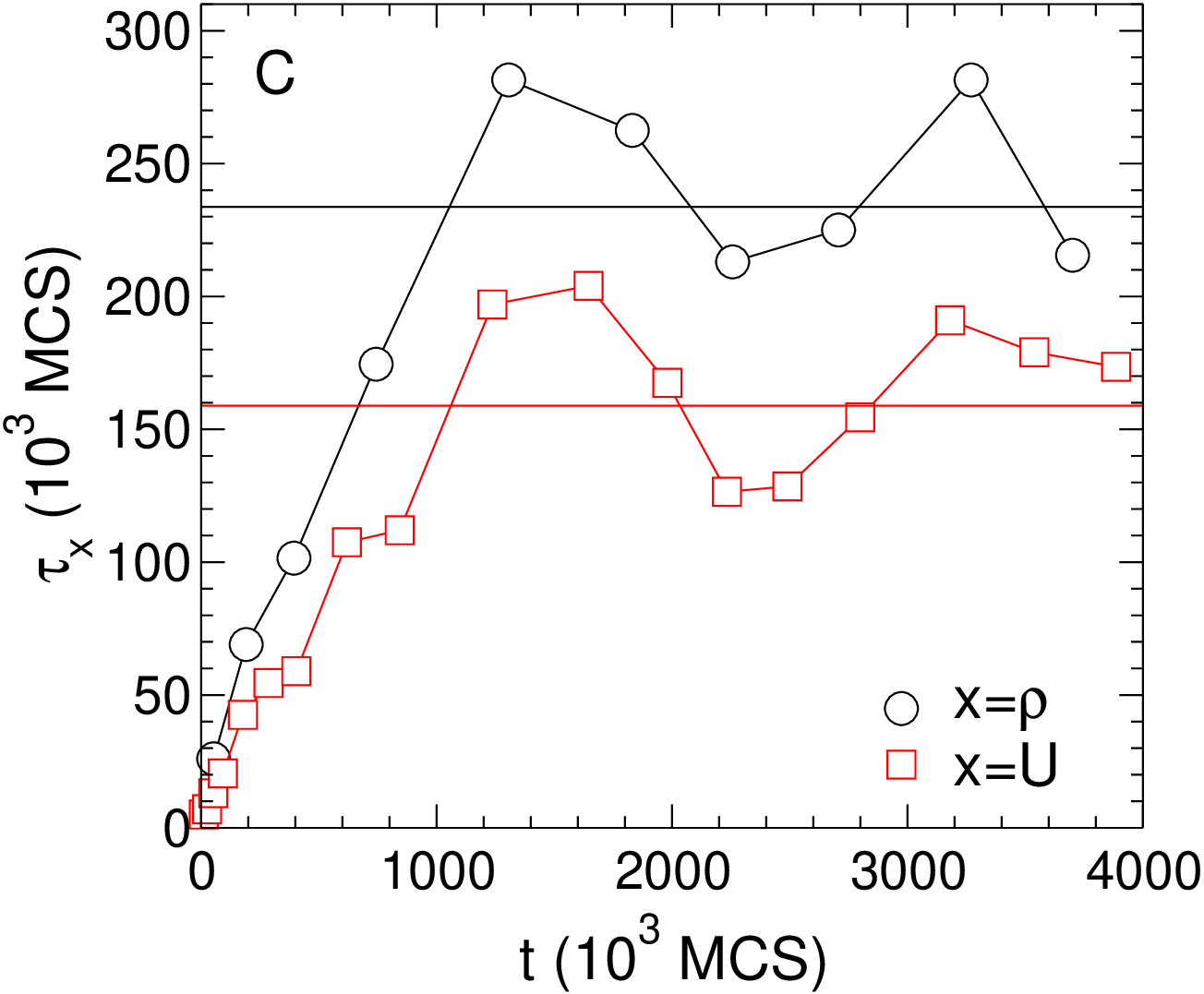}}
\caption{Relaxation times 
$\tau_\rho$ (black) and $\tau_U$ (red) 
for ST2 runs A, B, and C (panels top to bottom).  Horizontal lines indicate the values of 
$\bar \tau_\rho$ (black) and $\bar \tau_U$ (red).  Note that each value of $\tau_x$ is plotted at the value of $t$ corresponding to $t_0$ at the end of the relaxation cycle from which $\tau_x$ is computed.
}
\label{tau}
\end{figure}

\begin{figure}\bigskip\bigskip
\centerline{\includegraphics[scale=0.35]{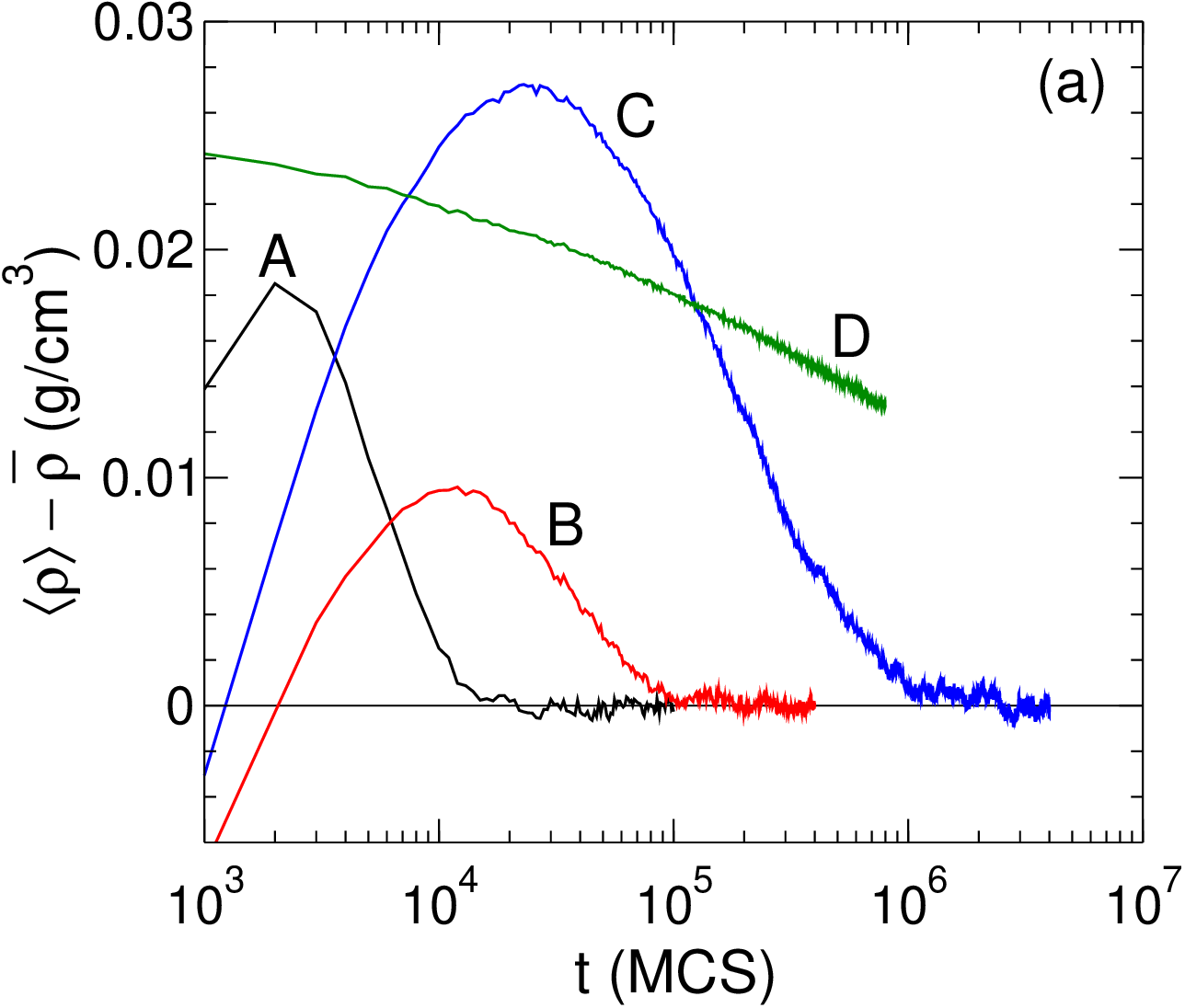}}\vskip 0.4 in
\centerline{\includegraphics[scale=0.35]{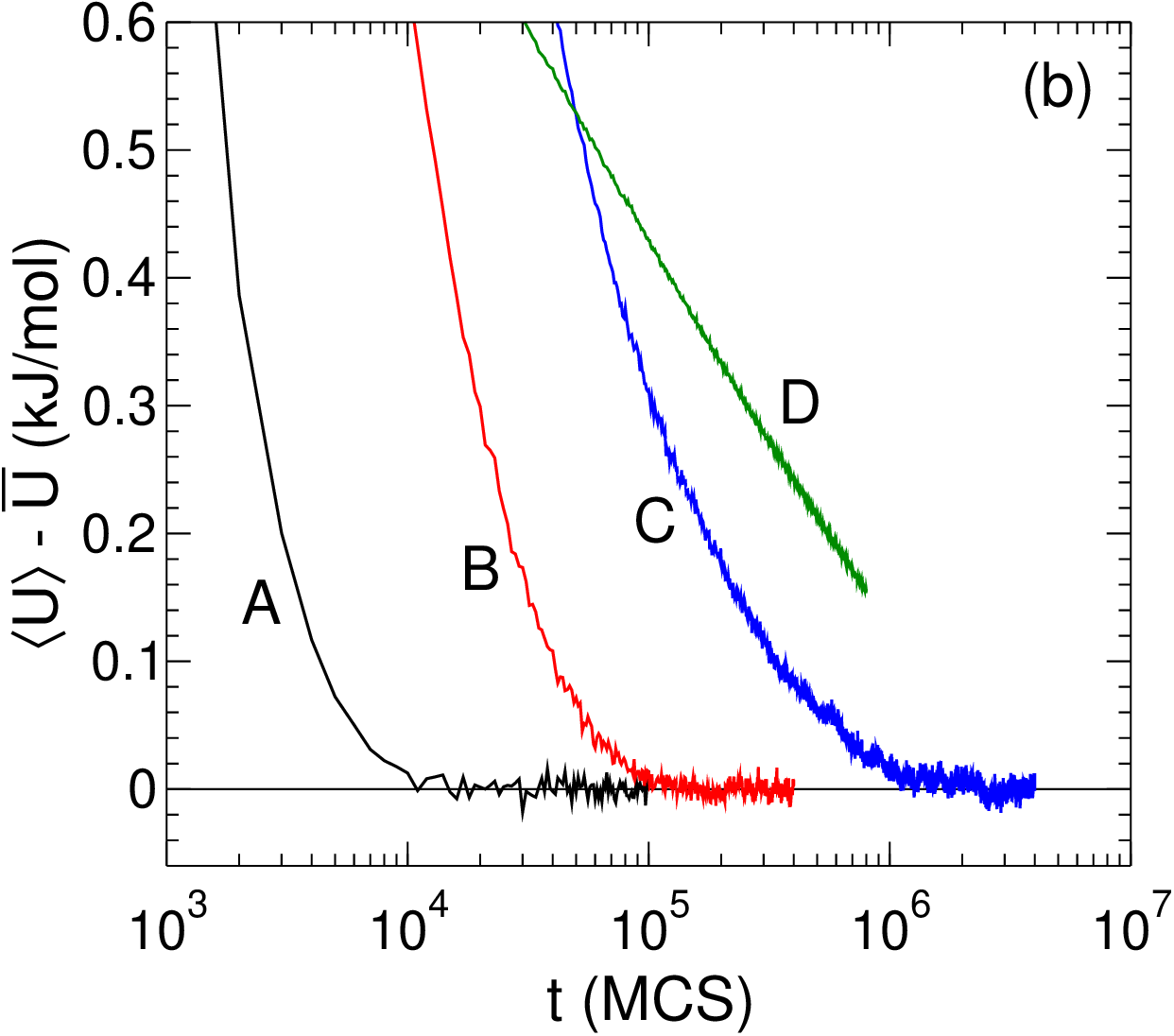}}
\caption{$\langle \rho \rangle$ (a) and $\langle U \rangle$
 (b) as a function of $t$ for ST2 runs A, B, C and D, plotted relative to the corresponding values of $\bar \rho$ or $\bar U$, and using a logarithmic time axis.
In the case of run D, $\bar \rho$ and $\bar U$ are not known.  To allow comparison with the other curves, for run D we arbitrarily set $\bar \rho=1.04$~g/cm$^3$ in (a) and $\bar U=-51.8$~kJ/mol in (b).}
\label{st2-all}
\end{figure}

\begin{figure}\bigskip\bigskip
\centerline{\includegraphics[scale=0.35]{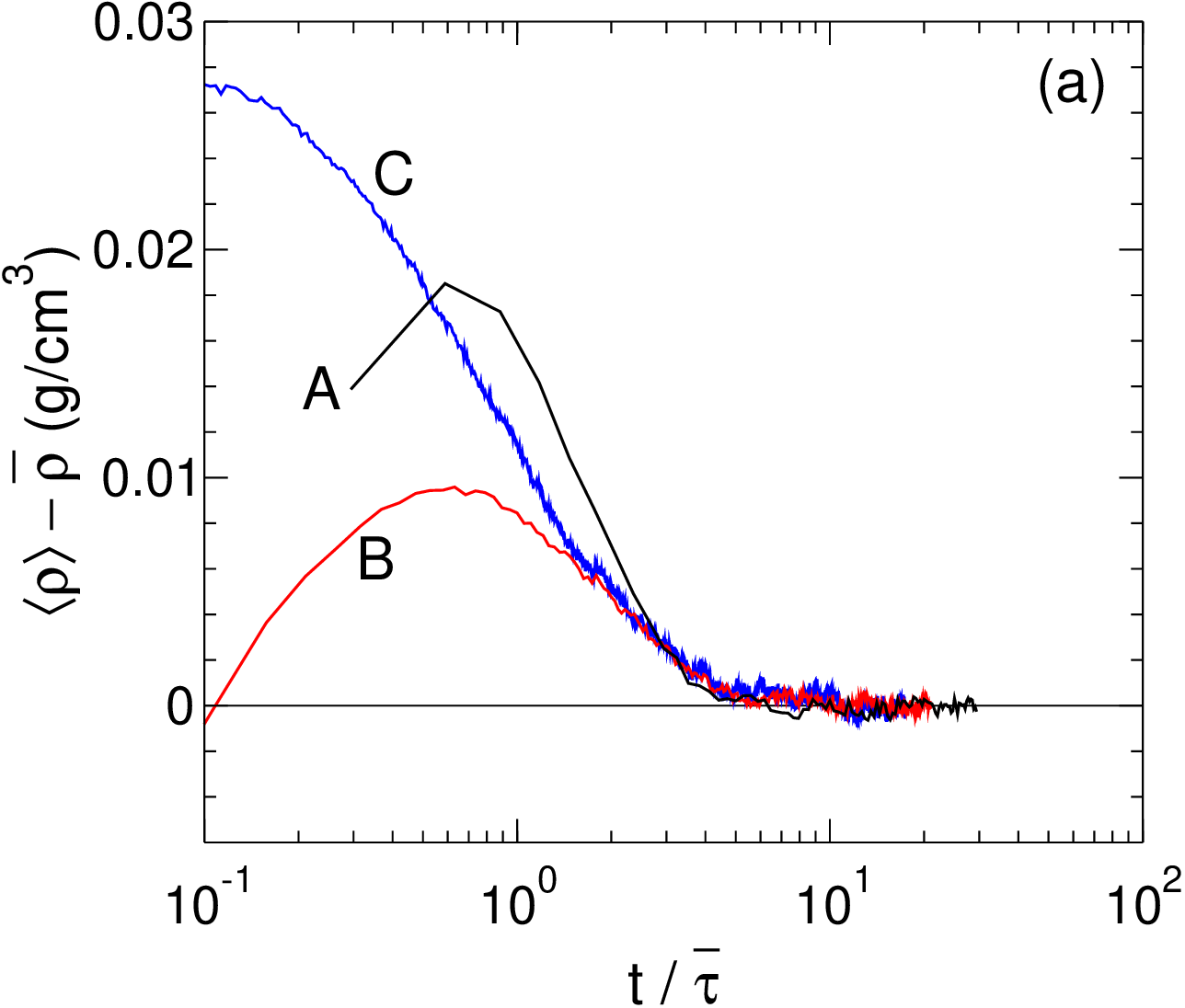}}\vskip 0.4 in
\centerline{\includegraphics[scale=0.35]{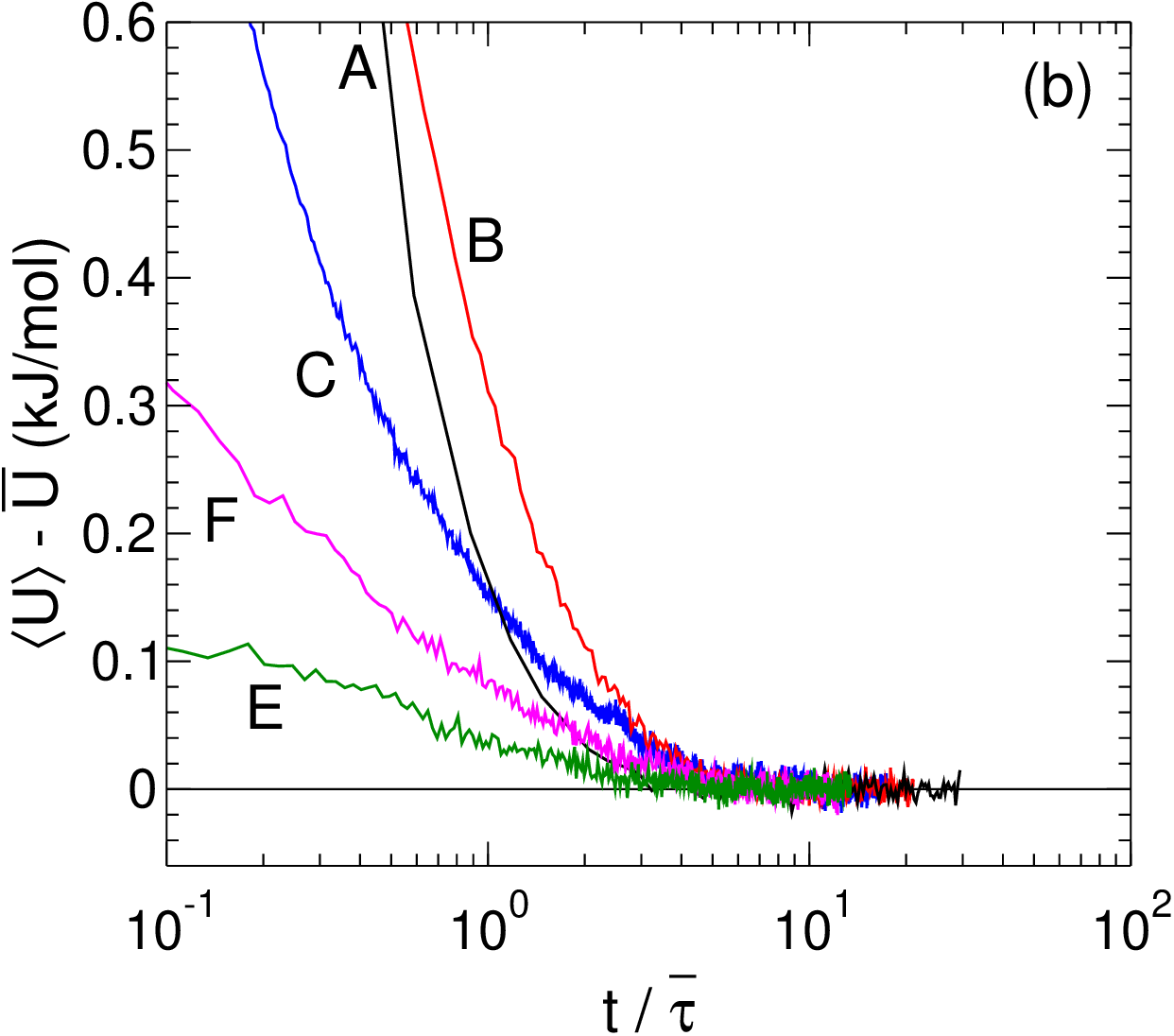}}
\caption{(a) $\langle \rho \rangle$ 
as a function of $t/\bar \tau$ for ST2 runs A, B and C, plotted relative to the corresponding value of $\bar \rho$.
(b) Same as in (a) but for $\langle U \rangle$, and comparing both our ST2 runs (A, B and C) and TIP4P/2005 runs (E and F).
}
\label{scale}
\end{figure}

\begin{figure}\bigskip\bigskip
\centerline{\includegraphics[scale=0.35]{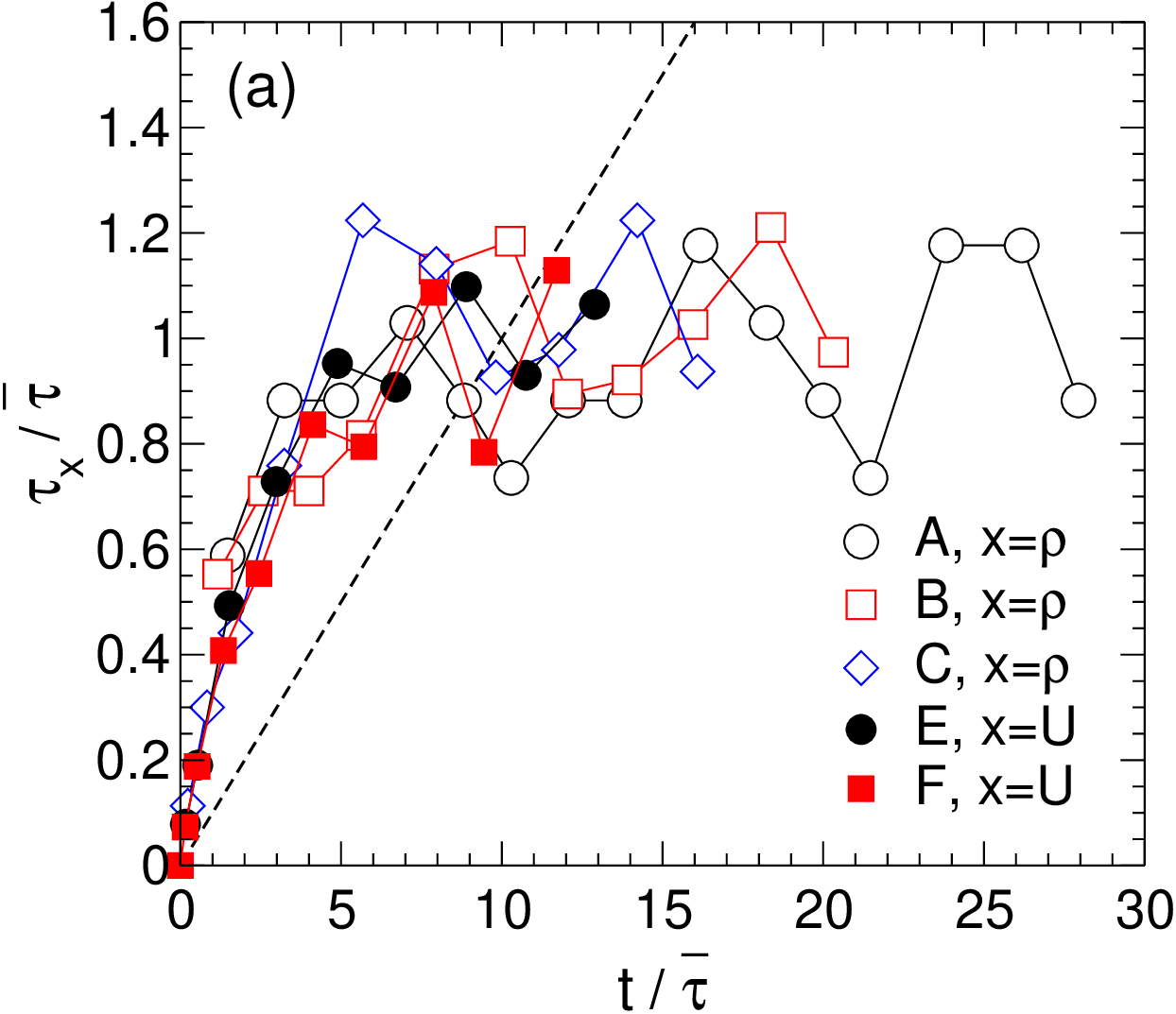}}\vskip 0.4 in
\centerline{\includegraphics[scale=0.35]{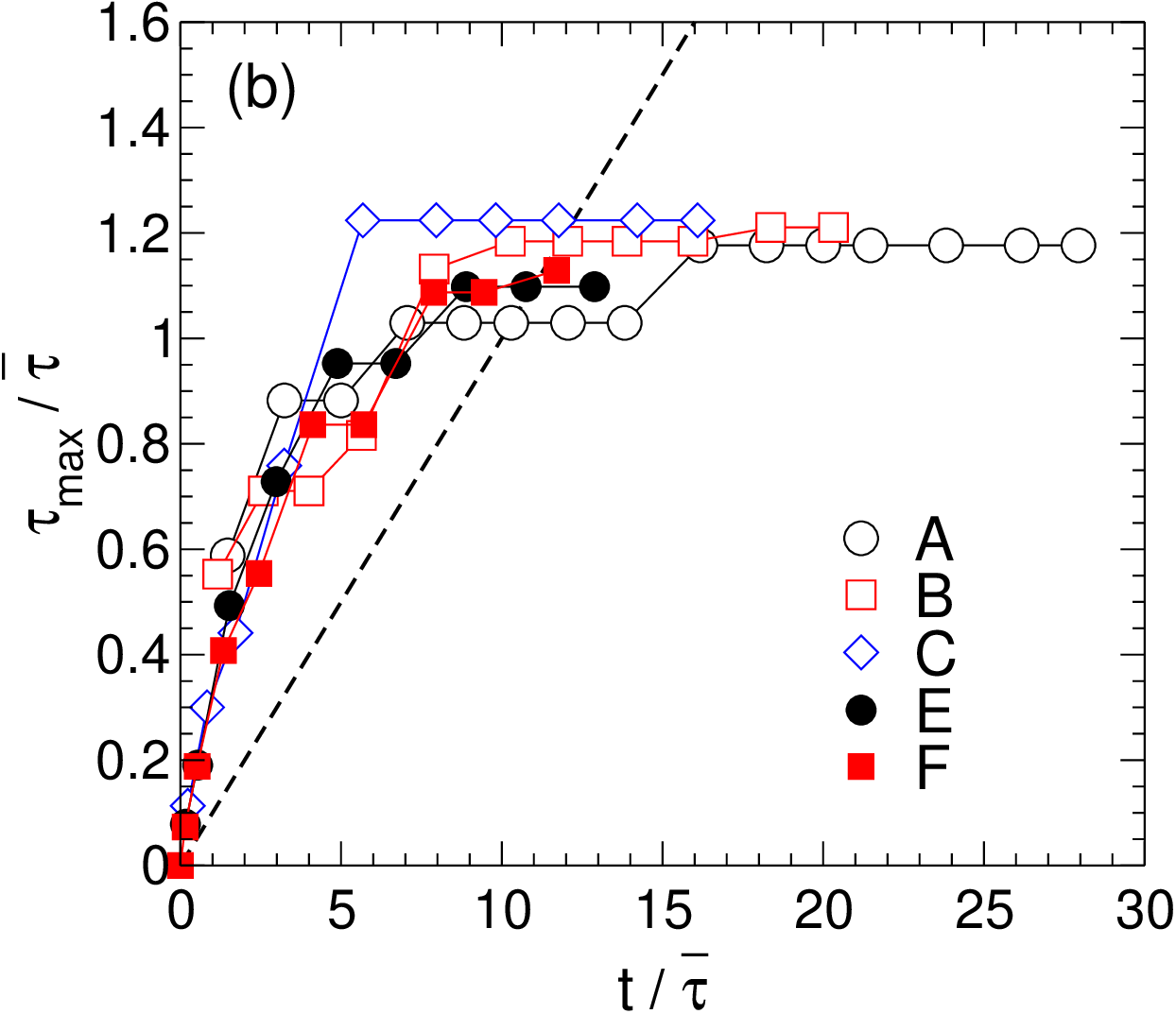}}
\caption{
(a) $\tau_x/\bar \tau$ and 
(b) $\tau_{\rm max}/\bar \tau$
versus $t/\bar \tau$, for 
runs A, B, C, E and F.
Note that each value of $\tau_x$ or $\tau_{\rm max}$
is plotted at the value of $t$ corresponding to $t_0$ at the end of the relaxation cycle from which it is computed.  The dashed line has slope $1/10$.
}
\label{tstop}
\end{figure}

\begin{figure}\bigskip\bigskip
\centerline{\includegraphics[scale=0.35]{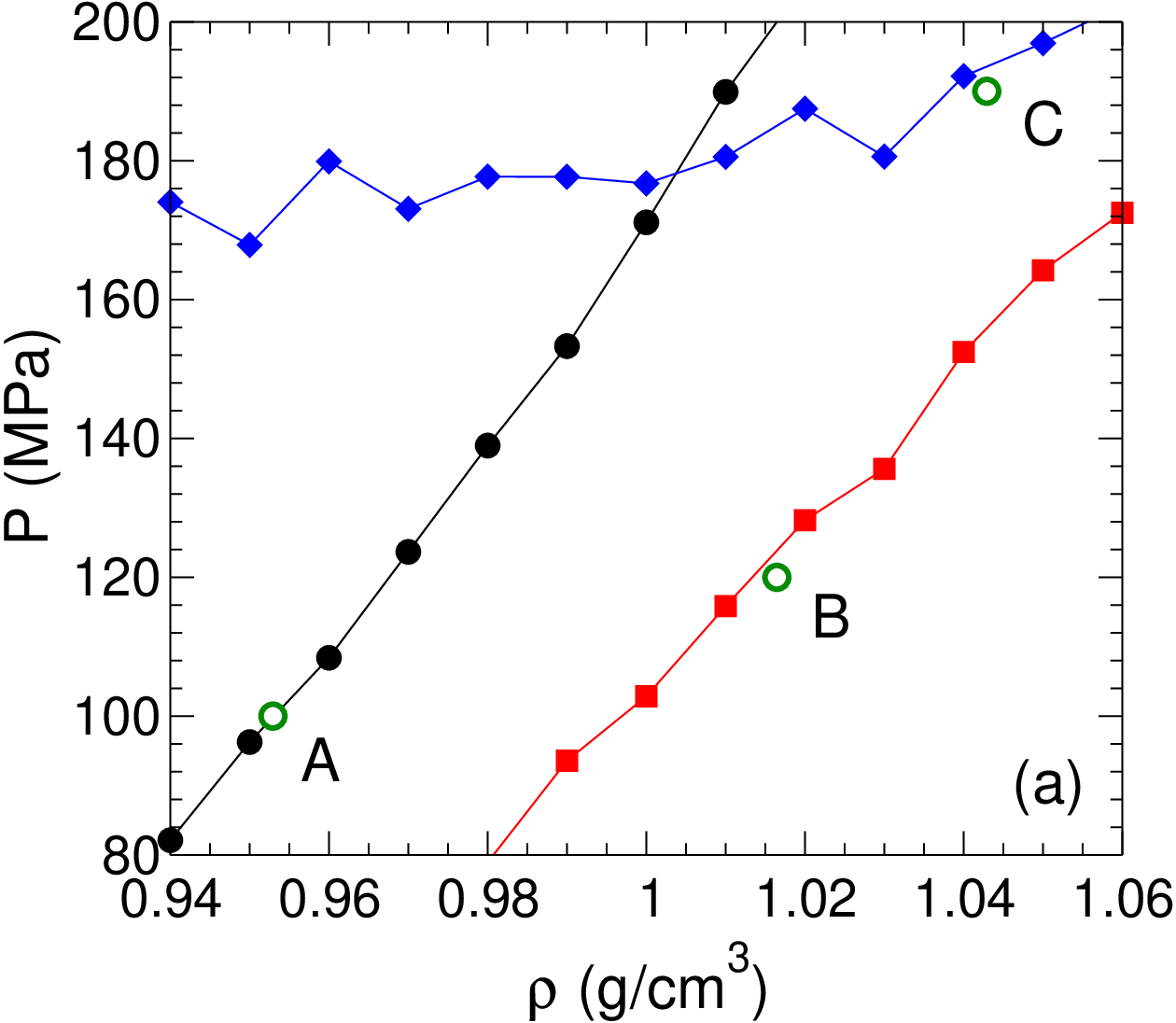}}\vskip 0.4 in
\centerline{\includegraphics[scale=0.35]{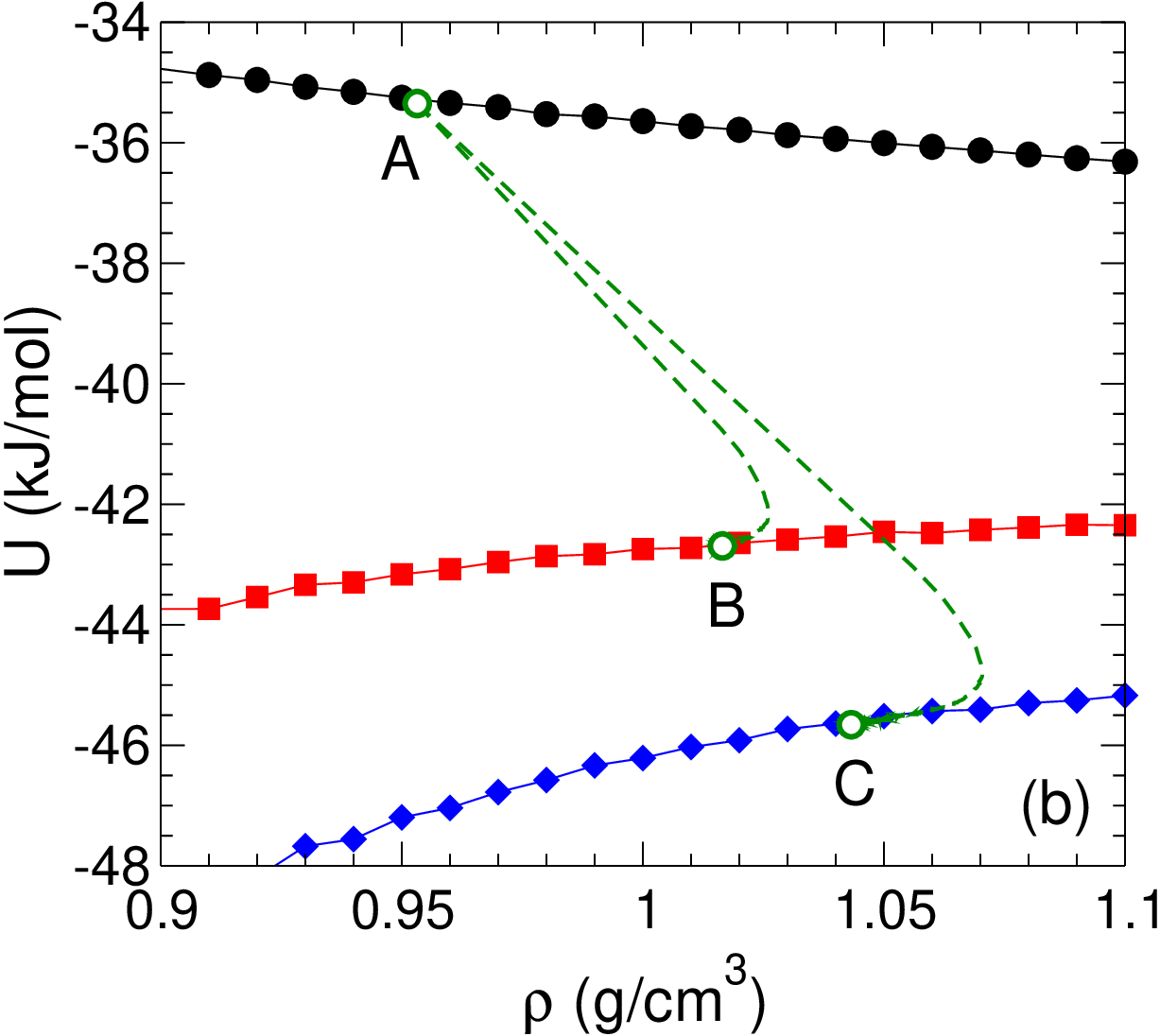}}
\caption{Comparison of our results for $\langle \rho \rangle$ and $\langle U \rangle$ for ST2 runs A, B and C
(open green circles) with ST2 data taken from Ref.~\cite{denmin} (filled symbols).  
The error for the green circles is smaller than the symbol size in both plots.  Panel (a) shows isotherms of $P$ versus $\rho$ from Ref.~\cite{denmin} for $T=400$~K (black), $290$~K (red), and $250$~K (blue).  Panel (b) shows isotherms of $U$ versus $\rho$ from Ref.~\cite{denmin} for the same $T$ as in (a).  In (b) we also show the parametric curves (green dashed lines) for $\langle \rho \rangle$ and $\langle U \rangle$ for runs B and C as they
evolve from their starting values at A to their equilibrium values.
}
\label{PD}
\end{figure}

\section{Definitions}
\label{defs}

Consider an ensemble of $M$ independent runs in which an observable $x(i,t)$ is measured in run $i$ of the ensemble as a function of time $t$.  In the following we use $\langle \cdots \rangle$ to denote an ensemble average over the runs at fixed $t$.  The ensemble average of $x$ over all runs at a fixed $t$ is defined as,
\begin{equation}
\big\langle x(t) \big\rangle = \frac{1}{M}\sum_{i=1}^{M} x(i,t).
\end{equation}
The variance of $x$ is,
\begin{equation}
\sigma_x^2(t)=\bigg\langle \Big[x(i,t)-\big\langle x(t)\big\rangle\Big]^2\bigg\rangle,
\end{equation}
where $\sigma_x$ is the standard deviation of $x$ at fixed $t$, which characterizes the average deviation of $x$ from $\langle x \rangle$ at time $t$.
Since $\langle x \rangle$ is an average of $M$ completely independent values of $x$, the standard deviation of the mean $s_x=\sigma_x/\sqrt{M}$ characterizes the error in our estimate of $\langle x \rangle$.

Following standard practice, we define the autocorrelation function for $x$, as measured from a reference time $t_0$, as,
\begin{equation}
C_x(t_0,t)= \frac {\bigg\langle \Big[x(i,t_0)-\big\langle x(t_0)\big\rangle\Big]\Big[x(i,t)-\big\langle x(t)\big\rangle\Big]\bigg\rangle}
                 {\sigma_x (t_0)\, \sigma_x(t)}.
\label{Ceq}
\end{equation}
As a function of the time difference $\Delta t=t-t_0$, $C_x$ measures the decay of the correlations between the fluctuations of $x$ from the ensemble average $\langle x \rangle$ occurring at $t$, and the fluctuations occurring at $t_0$.  We emphasize that only ensemble averaging is used in the definition of $C_x$.  Since our ensemble of runs is large, there is no need to average over different choices of the time origin $t_0$ in order to obtain an accurate value for $C_x$, as is commonly done when evaluating an autocorrelation function from a SLR.  This feature allows us to compute $C_x$ for any value of $t_0$ both during the approach to equilibrium, as well as after equilibrium has been established.  

As documented in the Appendix, it is straightforward to show that the standard deviation of fluctuations of $C_x$ as $C_x \to 0$ is exactly $M^{-1/2}$.  This result is important in the present context because it establishes how large $M$ must be in order to effectively use $C_x$ to monitor the relaxation of the ensemble of runs to equilibrium.  If we choose $M=1000$, then
$1/\sqrt{M}=0.032$, and so when $C_x$ approaches zero, it will do so with fluctuations that remain within $\pm 2/\sqrt{M}=\pm 0.064$ of zero for $95\%$ of the time.  As we will see below, this error is sufficiently small to allow for the accurate evaluation of the relaxation time for the system, starting from any given $t_0$.

\section{Test cases}

\subsection{Bulk ST2 water}

Our first test case is a Monte Carlo simulation of bulk water, using the ST2 intermolecular potential.  We employ the ST2 model of water in the original form proposed by Stillinger and Rahman~\cite{st2}, using the reaction field method to approximate the long-range contribution of the electrostatic interactions~\cite{rf}.  ST2 water has been extensively studied in previous work, mainly to investigate the liquid-liquid phase transition that occurs in the supercooled region of the phase diagram for this model.  As a consequence, there is a rich literature of published work to which we can compare our results~\cite{PSES,denmin,becker,holten,palmer2014}.  The ST2 simulations presented here are part of a larger study of ice nucleation in supercooled water, to be published separately~\cite{siobhan}.

Our Monte Carlo simulations of ST2 water are carried out in the constant-$(N,P,T)$ ensemble, with $N=1728$ molecules contained in a cubic simulation cell, with periodic boundary conditions.  One Monte Carlo step (MCS) consists of (on average) of $N-1$ attempted rototranslational moves, and one attempted change of the system volume. The maximum size of the attempted rototranslational and volume changes are chosen to give MC acceptance ratios in the range 30\% to 40\%.

To initialize a swarm of independent runs, we generate $M=1000$ different configurations, each of which consists of $N$ water molecules with their centers of mass arranged on a simple cubic lattice of density $\rho=1.0$~g/cm$^3$, and with randomized molecular orientations.  These configurations are used to initialize a swarm of runs at $T=400$~K and $P=100$~MPa (labelled run A in Table~\ref{table}).  
Each run in this swarm is carried out for a run time of $t_{\rm run} =10^5$~MCS.

As summarized in Table~\ref{table}, 
the final configurations generated in run A are used to initialize two new swarm runs, B and C. Run B aims to characterize a state point on the ice-liquid coexistence line for ST2 water, and run C studies a state point close to the liquid-liquid critical point of ST2 water.  The final configurations of run C are then used to initialize a swarm of runs D, which studies a low temperature state at which the system is quenched into a glass, and where (as we will see) the system is unable to achieve liquid-like equilibrium on the time scale currently accessible to simulations.  Table~\ref{table} gives the values of $T$, $P$, and $t_{\rm run}$ for each of these ST2 swarm runs.

Fig.~\ref{st2} 
shows the time dependence of the ensemble-averaged density $\langle \rho \rangle$ 
and the potential energy $\langle U \rangle$ 
for swarm runs A, B, and C.  
In each case, $t=0$ corresponds to the set of microstates used to initialize the ensemble, as indicated in Table~\ref{table}.
For runs A, B and C, $t_{\rm run}$ is sufficiently large that the time dependence of 
$\langle \rho \rangle$ and $\langle U \rangle$ 
in Fig.~\ref{st2} 
suggests that an approximately steady state has been attained for $t>t_{\rm run}/2$, if not earlier.  For each ensemble, we evaluate the time average of $\langle \rho \rangle$ and $\langle U \rangle$ 
for $t_{\rm run}/2<t<t_{\rm run}$, 
respectively denoted $\bar \rho$ and $\bar U$.  
In each panel of Fig.~\ref{st2}, the horizontal solid line passing through the middle of the data at large $t$ identifies the corresponding value of $\bar \rho$ or $\bar U$.  The horizontal lines that bracket $\bar \rho$ and $\bar U$ identify values at $\bar\rho \pm 2\bar s_\rho$ and $\bar U\pm 2\bar s_U$ respectively, where $\bar s_\rho$ and $\bar s_U$ are time averages of $s_\rho$ and $s_U$ for $t_{\rm run}/2<t<t_{\rm run}$.
We see in Fig.~\ref{st2} that the fluctuations of $\langle \rho \rangle$ and $\langle U \rangle$ 
are largely confined to the ranges $\bar \rho \pm 2\bar s_\rho$ and $\bar U \pm 2\bar s_U$ in the second half of each run.  This behavior is consistent with $\langle \rho \rangle$ and $\langle U \rangle$ having reached equilibrium, since in this case we would expect them to fluctuate within a range of $\pm2\bar s_x$ for $95\%$ of the time.

Fig.~\ref{C} 
shows $C_\rho$ and $C_U$ 
for runs A, B and C evaluated as a function of $\Delta t$ for $t_0=t_{\rm run}/2$, a time by which equilibrium has been established according to the results presented in Fig.~\ref{st2}.  The time scale for the decay of $C_\rho$ and $C_U$ to zero therefore reflects the equilibrium relaxation time of each state point.  We find in each case that $C_\rho$ and $C_U$ decay to zero in a time that is shorter than $t_{\rm run}/2$, confirming that our runs are able to relax completely within equilibrium.
The dotted horizontal lines in Fig.~\ref{C} locate $\pm 2 M^{-1/2}$.  We find that the fluctuations of $C_x$ as $C_x\to 0$ are largely confined within these bounds, as predicted in Section~\ref{defs}.

We also evaluate $C_\rho$ and $C_U$ 
for various values of $t_0$, shown as the blue ``saw-tooth" curves in Fig.~\ref{st2}.
These curves are calculated as follows:
Starting at $t_0=10^3$~MCS, we evaluate the decay of $C_x$ as a function of $t$, for both $x=\rho$ and $x=U$.  
At the next smallest time such that $C_x<e^{-2}$, we reset $t_0$ to the current time, and continue evaluating $C_x$.  This process is repeated for the duration of the run, thus generating a saw-tooth curve that quantifies 
successive cycles of relaxation, both as the ensemble evolves towards equilibrium, and after equilibrium has been established.  

As shown in Fig.~\ref{st2}, 
we find that the decay of $C_x$ is approximately exponential (i.e. $\log C_x$ is linear in $t$), especially in the case of $C_\rho$.  We therefore define the relaxation time $\tau_x$ as $1/2$ of the time required for $C_x$ to first reach $e^{-2}$ during each relaxation cycle.  Fig.~\ref{tau} 
shows $\tau_\rho$ and $\tau_U$
as a function of $t$ for runs A, B and C.  Consistent with Fig.~\ref{st2}, Fig.~\ref{tau} shows that $\tau_x$ is approximately constant in the 2nd half of our runs.  
{\color{black} We note that $\tau_x$ initially increases with $t$ before reaching a steady state.  This is to be expected for runs B and C in part because the initial configurations come from runs at higher $T$, where the equilibrium relaxation time is shorter.  Also, in all cases, the system is far out of equilibrium at the beginning of the runs, providing a strong initial driving force for change, demonstrated by the rapid decay of the autocorrelation functions at early times.}

To characterize the average equilibrium relaxation time $\bar \tau$ for each state point, we first compute $\bar \tau_\rho$ and  $\bar \tau_U$,
the average values of $\tau_\rho$ and $\tau_U$ for $t_0>t_{\rm run}/2$.  We then define $\bar \tau=\max\{\bar \tau_\rho,\bar \tau_U\}$, to ensure that we use the most conservative choice of the relaxation time available.
The values for $\bar \tau$ 
so obtained are given in Table~\ref{table}.  We note in all cases that $\bar \tau_\rho$ is greater than $\bar \tau_U$.

In each panel of Fig.~\ref{st2}, the open red symbols present values of $\langle \rho \rangle$ and $\langle U \rangle$ 
at the values of $t_0$ that mark the beginning of a new relaxation cycle in the saw-tooth curve for $C_x$.  The error bars on each data point represent $\pm 2s_\rho$ and $\pm 2s_U$ respectively, the instantaneously calculated error in 
$\langle \rho \rangle$ and $\langle U \rangle$. 
These data demonstrate that the error in 
$\langle \rho \rangle$ and $\langle U \rangle$ does not vary significantly with $t$ during the evolution of the swarm to equilibrium.  These data also show that the instantaneous values of $\langle \rho \rangle$ and $\langle U \rangle$ attain values that are within error of $\bar \rho$ and $\bar U$ well before $t_{\rm run}/2$.

Fig.~\ref{st2-all} 
shows $\langle \rho \rangle$ and $\langle U \rangle$ 
plotted with a logarithmic time axis.  
The time dependence of $\langle \rho \rangle$
exhibits a non-monotonic approach to the
equilibrium value, possibly arising from the time separation between the vibrational and configurational degrees of freedom~\cite{kovacs1963sci,mossa2004crossover}.
Fig.~\ref{st2-all} 
also confirms that a stable equilibrium has been attained at large $t$ for runs A, B, and C.  Fig.~\ref{scale} 
shows the time dependence of $\langle \rho \rangle$ and $\langle U \rangle$, where the time has been scaled by $\bar \tau$.  Fig.~\ref{tstop}(a) 
shows a similar plot for the time dependence of $\tau_\rho$.
Figs.~\ref{scale}  and~\ref{tstop}(a) 
demonstrate that in all cases, equilibrium thermodynamic properties and equilibrium relaxation times are established on a time scale of $10 \bar \tau$ or less.

To test if the present results agree with previously reported results for ST2 water, Fig.~\ref{PD} 
compares our results for $\langle \rho \rangle$ and $\langle U \rangle$ 
from runs A, B and C with results for ST2 water based on the data set generated for Ref.~\cite{denmin}.  
The data reported in Ref.~\cite{denmin} was obtained from constant-$(N,V,T)$ molecular dynamics simulations with $N=1728$.  To conduct this comparison, we use the values of $\langle \rho \rangle$ and $\langle U \rangle$ evaluated at $t=t_{\rm run}$ (green open symbols in 
Fig.~\ref{st2}).
The agreement between the two data sets is excellent, and again confirms that we have obtained equilibrium properties using our swarm relaxation strategy.  Note in Fig.~\ref{PD} that the error for our data points ($\pm 2 s_x$) is much smaller than the symbol size.  The scatter in the data points taken from Ref.~\cite{denmin} is larger, indicating that the estimates obtained here are of higher precision that those reported in Ref.~\cite{denmin}.  

In the case of run D, as expected, the swarm does not reach equilibrium on the time scale of our simulations.  In Fig.~\ref{C} we see that both $C_\rho$ and $C_U$ remain very far from zero throughout the simulation time.  Fig.~\ref{st2-all} shows that both $\langle \rho \rangle$ and $\langle U \rangle$ 
continue to vary with $t$ even at the largest $t$.  It is apparent that a much longer simulation would be required to bring run D into equilibrium.  
Our results from run D confirm that the swarm relaxation strategy used here is able to clearly distinguish between a liquid and a glassy state.

\subsection{TIP4P/2005 water nanodroplet}

As a second test case, we present molecular dynamics simulations of an isolated nanodroplet of $N=360$ water molecules, surrounded by vacuum.  In this case, the water interactions are modelled using the TIP4P/2005 potential~\cite{tip4p}.  These simulations are also used in a study of water nanodroplets over a wide range of $N$ and $T$~\cite{m1}.  In the present simulations, we focus on $T=180$~K, where $T$ is controlled using a Nose-Hoover thermostat~\cite{nose,hoover}.  We use a cubic simulation cell of linear dimension $L=$10~nm, with periodic boundary conditions.  The liquid nanodroplet occupies less than $2\%$ of the total volume of the simulation cell.  Since the diameter of the nanodroplet is significantly smaller than $L/2$, we directly evaluate all electrostatic interactions among molecules separated by a distance of less than $L/2$, and ignore interactions beyond this distance.

\begin{figure}\bigskip\bigskip
\centerline{\includegraphics[scale=0.35]{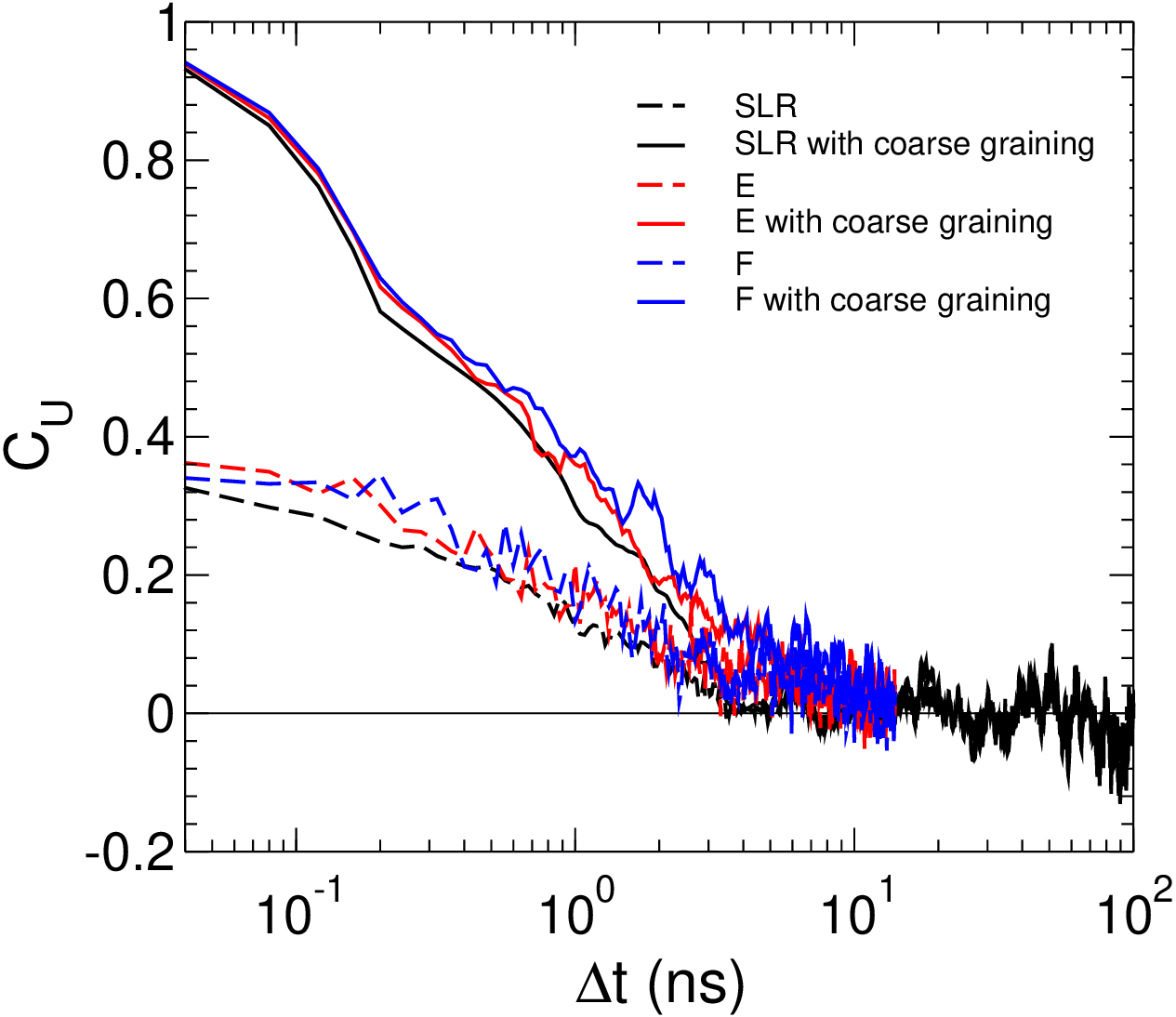}}
\caption{Comparison of $C_U$ from our TIP4P/2005 nanocluster runs.  Shown are $C_U$ for the SLR (black), and the swarm runs E (red) and F (blue).  Note that $\Delta t=t-t_0$.  For runs E and F, we choose $t_0=10$~ns.  Results for $C_U$ both with (solid) and without (dashed) coarse graining are shown.
}
\label{sC}
\end{figure}

First we conduct a SLR of this $180$~K nanodroplet lasting 2700~ns, to compare to our swarm runs.  
The initial configuration for this SLR is an equilibrium configuration taken from a single long nanodroplet simulation conducted at $250$~K.
The potential energy $U$ is recorded every 40~ps during the SLR at $180$~K.  From the time series for $U$ over the last 288~ns of the SLR, we evaluate the autocorrelation function using the definition in Eq.~\ref{Ceq}, but where the ensemble average is replaced by an average over the choice of time origin $t_0$.  This autocorrelation function, plotted in Fig.~\ref{sC}, 
exhibits a fast initial decay, due to large fluctuations which occur on a time scale of less than 40~ps, followed by a slower relaxation to zero.  Since it is the slower relaxation to zero that we wish to characterize, we coarse grain the time series by averaging our data for $U$ over successive, non-overlapping time windows of 200~ps.  The autocorrelation function for the coarse grained time series is also shown in Fig.~\ref{sC}.  As desired, the coarse grained time series yields an autocorrelation function that better spans the full range of decay from 1 to 0 within the time domain studied here.  Fig.~\ref{sC} shows that the relaxation time $\bar\tau$ for our SLR is on the order of 1~ns, confirming that this run is long enough for measuring equilibrium properties. 

\begin{figure}\bigskip\bigskip
\centerline{\includegraphics[scale=0.35]{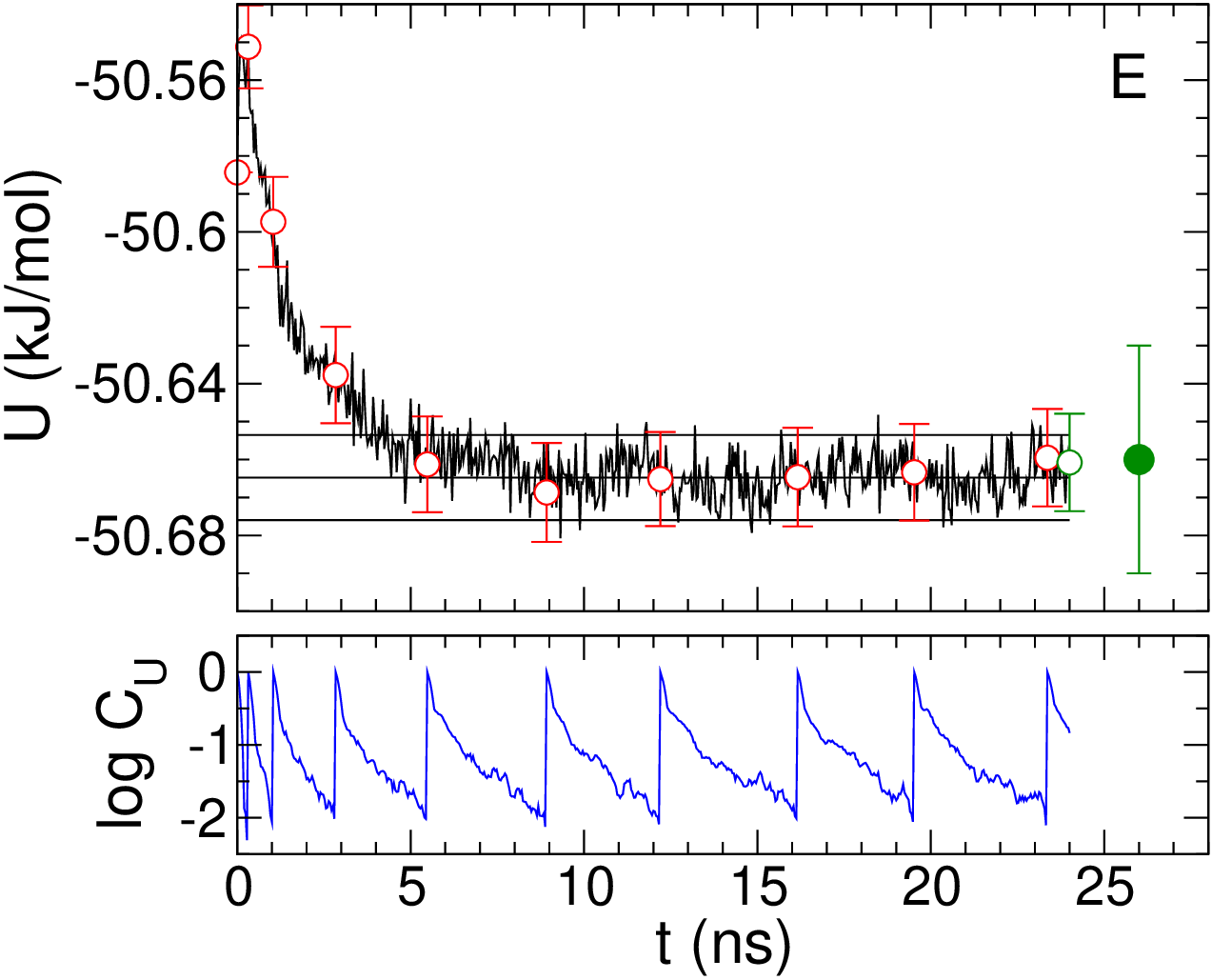}}\vskip 0.4 in
\centerline{\includegraphics[scale=0.35]{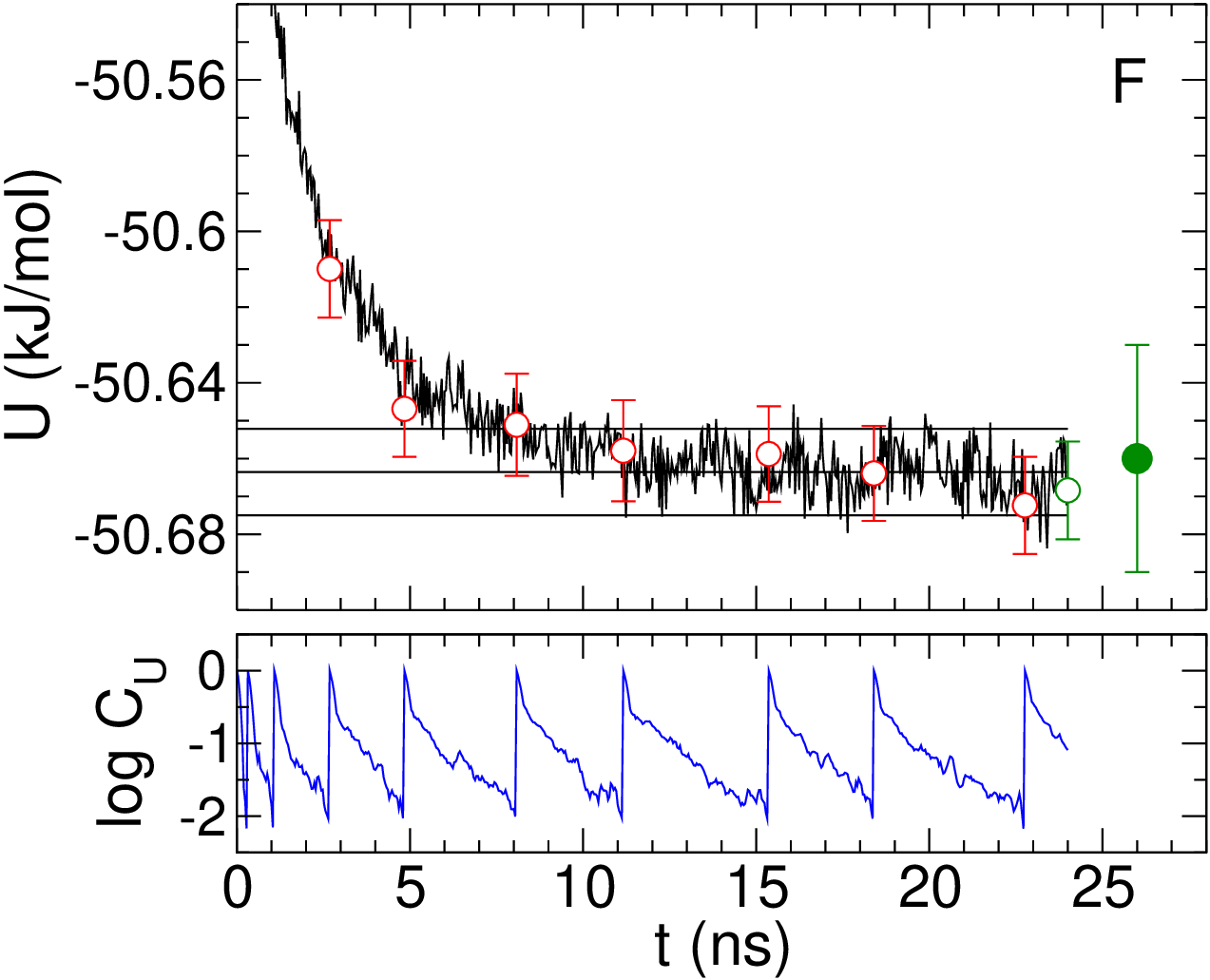}}
\caption{Time dependence of $\langle U \rangle$ 
(black curve) for TIP4P/2005 runs E and F (panels top to bottom).
The black horizontal lines identify $\bar U$ and $\bar U \pm 2 \bar s_U$.  
The lower section of each panel shows
$\log C_U$ (evaluated from the coarse grained times series for $\langle U \rangle$) 
over successive relaxation cycles, as described in the text.  
The red circles are values of $\langle U \rangle$ (with error $\pm 2 s_U$) at the beginning of each relaxation cycle.  
The green open circle is $\langle U \rangle$ at $t=t_{\rm run}$.
The green filled circle (displayed arbitrarily at $t=26$~ns) is $\bar U$ from our SLR, evaluated with error as described in the text.
}
\label{sU}
\end{figure}

We then conduct two swarm relaxation runs of the $N=360$ TIP4P/2005 water nanocluster, labelled E and F in Table~\ref{table}.  To initialize run E, we select one equilibrium configuration from the SLR conducted at $180$~K, and generate $M=1000$ copies, 
where we use the same spatial coordinates for the molecules in the system, but select their velocities (both translational and rotational) randomly from a Maxwell-Boltzmann distribution appropriate for $T=180$~K.  To initialize run F, we proceed in the same way as for run E, except that the initial configuration is an equilibrium configuration obtained from a separate SLR conducted at $T=220$~K.
We choose this approach to test the use of an ``isoconfigurational"~\cite{widmer2007study}
set of microstates to initialize a swarm relaxation run.  We anticipate that there may be many situations where a single configuration of a complex system is available, either near or away from the state we wish to equilibrate.  In this case, an isoconfigurational set is a convenient way to initialize a swarm relaxation run, compared to generating $M$ independent configurations from scratch.  Also, runs E and F will allow us to compare the time to recover the ensemble-average properties at $T=180$~K when starting from a single equilibrium microstate (E), versus an out-of-equilibrium microstate (F).

For runs E and F, our swarm relaxation simulations run for $t_{\rm run}=24$~ns.  
Fig.~\ref{sU} 
shows the time dependence of $\langle U \rangle$ 
obtained for E and F, as well as the successive relaxation cycles of $C_U$.  
Fig.~\ref{sU} demonstrates that $\langle U \rangle$ for both E and F is in a steady state when $t>t_{\rm run}/2$.
We note that despite the fact that the initial configuration used for run E is from the equilibrium portion of a SLR at the same conditions, its value of $U$ is well outside, and above, the error estimate for $\bar U$.  This occurs because $\sigma_U$ is much greater than $s_U$, and so it is likely that a randomly chosen single configuration from equilibrium will fall outside of 
$\bar U \pm 2 \bar s_U$.  
We also note that when calculating $C_U$ from our swarm runs, we coarse grain the time series of $U$ values for each run in the same way as described above for the SLR.  The values of $\tau$ obtained from the successive relaxation cycles are shown as the solid symbols in Fig.~\ref{tstop}(a), 
where we have used  $\bar\tau=1.8$~ns 
for run E, and $\bar\tau=1.9$~ns for run F, 
evaluated by averaging the values of $\tau$ from runs E and F for $t_{\rm run}/2 < t < t_{\rm run}$.  

In Fig.~\ref{sC} we plot $C_U$ as obtained from runs E and F, when $t_0=10$~ns, and using both the original and coarse grained time series for $U$.  We find that the autocorrelation functions obtained from the SLR for $T=180$~K and from the swarm runs E and F agree within error, confirming that the equilibrium relaxation time $\bar\tau$ is the same in all cases.

By time averaging over the last 288~ns of the SLR at $T=180$~K, we obtain $\bar U=-50.66\pm0.03$.   Here the error has been evaluated as $2\sigma/\sqrt{N_{\tau}}$, where $\sigma$ is the standard deviation of the time series for $U$, and $N_{\tau}=(288\,{\rm ns})/\bar\tau$, with $\bar\tau=1.8$~ns.  That is, we have made the (optimistic) assumption that successive independent configurations are separated by $\bar\tau$ in the SLR.  This value of $\bar U$ is plotted as the solid green circle in both panels of Fig.~\ref{sU}, which demonstrates that the equilibrium value of $U$ obtained from the swarm runs E and F, and the SLR, all agree within error.  
Our results also show that the equilibrium values of $\tau$ (Fig.~\ref{tstop}) and $\langle U \rangle$ [Fig.~\ref{scale}(b)]
in runs E and F are established within a run time of $10\bar\tau$, regardless of whether our swarm runs are initiated from an equilibrium or out-of-equilibrium configuration.

\section{Computational efficiency}

The above results indicate that all investigated test cases attain equilibrium within a time less than $10\bar\tau$.
This time scale is physically reasonable:  When the equilibrium we seek to attain is more slowly relaxing than the state point from which our swarms are launched, it is not surprising that the time scale to reach equilibrium is dominated by the time scale for relaxation within equilibrium.  We define $\bar\tau$ as the time to relax an equilibrium autocorrelation function to $1/e$, and so full decorrelation requires several $\bar\tau$; e.g. $5\bar\tau$ is required for an exponential autocorrelation function to decay to less than 0.01, and longer would be required for a stretched exponential.  Hence our test swarms reach equilibrium in less than two full decorrelation times of the equilibrium system.

For our swarm relaxation strategy to be both accurate and efficient, the runs need to be stopped at a time $t_{\rm stop}$ 
that is longer than the time required for the system to attain equilibrium, but not much longer.
That is, if each run only contributes one microstate to the ensemble averages, then continuing the runs in the equilibrium time regime is a waste of computing resources.  Based on the results shown above, 
$t_{\rm stop}=10\bar \tau$ 
would be a good choice, but $\bar \tau$ 
is not known in {\it a priori}.  However, a reliable estimate for $t_{\rm stop}$ 
can still be made due to the fact that we can monitor $\tau$ as a function of $t$ during the simulations.  In the present context, by $\tau$ we mean the time-dependent relaxation time for the most slowly relaxing observable of interest.

In particular, it is reasonable to assume that the approach of
$\tau$ to $\bar \tau$ (from below) is approximately exponential in $t$.  If we also assume that $\tau=\bar \tau$ for $t>10\bar \tau$, then the function $\tau(t)$ will lie above the linear curve $t/10$ from $t=0$ to some time $t\le 10\bar \tau$, and will lie below $t/10$ for $t>10\bar \tau$.  The time at which the curves for $\tau(t)$ and $t/10$ cross thus provides a way to estimate (an upper bound on) $\bar \tau$.  We see in Fig.~\ref{tstop}(a)
that such a crossing is observed in each case studied here.  

We therefore propose the following procedure to determine
$t_{\rm stop}$:
Let $\tau_{\rm max}(t)$ 
be the largest value of $\tau$ observed so far in a swarm run of length $t$.  We define $t_{\rm stop}$ as the smallest $t$ satisfying $t>10\tau_{\rm max}(t)$.  This procedure allows $t_{\rm stop}$ to be identified using only information that is available at time $t$.  We use $\tau_{\rm max}(t)$ instead of $\tau(t)$ in order to make the estimate of $t_{\rm stop}$ a conservative one.  In Fig.~\ref{tstop}(b) we plot $\tau_{\rm max}(t)$ 
for each state point, from which we obtain estimates for $t_{\rm stop}$ from the crossing time of the curves for $\tau_{\rm max}(t)$ and $t/10$.  These values of $t_{\rm stop}$ 
are tabulated in Table~\ref{table}.  In all cases, we find $t_{\rm stop}$ is larger than $10 \bar \tau$, but not too much larger; 
$t_{\rm stop}/\bar \tau$  ranges between $12$ and $14$.

Next, we compare the efficiency of our swarm relaxation strategy relative to a SLR.
Let us denote the time separation between independent microstates during a run as $n\bar \tau$, leaving open for the moment what a good choice of $n$ should be.
A SLR that generates $K$ independent microstates will run for a wall-clock time of 
$t_{\rm SLR}=n\bar \tau K$.  Here we ignore the equilibration time of a SLR, by assuming that this is a small fraction of the total run length.
Using the swarm relaxation approach, and the procedure described above to determine $t_{\rm stop}$, each run will terminate after approximately $13\bar \tau$. 
Using $M$ processors concurrently, subject to the constraint $M\le K$, the swarm approach will generate $K$ independent microstates in a wall-clock time of $t_{\rm swarm}=13\bar\tau K/M$. 
The swarm strategy is thus faster, in terms of wall-clock time, than a SLR by a speedup factor of 
$f_{\rm speedup}=t_{\rm SLR}/t_{\rm swarm}=nM/13$.  
The total computational cost for a swarm run relative to a SLR increases by a factor of 
$f_{\rm cost}=Mt_{\rm swarm}/t_{\rm SLR}=13/n$.  

As for the choice of $n$, many simulation studies consider microstates to be independent if they are separated by as little as $\bar\tau$; see e.g. Ref.~\cite{poole2013}.  However, this choice almost certainly underestimates the error in a SLR, relative to the error evaluated in a swarm run.  As discussed above, complete decorrelation requires several $\bar\tau$, e.g. $n=5$.  Since the swarm approach produces completely independent microstates, for a direct comparison we should consider a SLR from which only completely independent microstates have been harvested.  Hence, for comparing the two approaches, we choose $n=5$.  
As shown above, a practical value for both $K$ and $M$ is 1000.  With these choices, the swarm approach is faster than a SLR by a factor of $f_{\rm speedup}=385$, in exchange for a total computational cost that increases by a factor of $f_{\rm cost}=2.6$.

The above estimates for $f_{\rm speedup}$ and $f_{\rm cost}$
are approximate, and can be expected to vary substantially for different systems, different parameter choices (such as for $n$), and as the strategy for implementing a swarm approach is varied to best suit a particular physical system and/or computing facility.  Although our results are thus difficult to generalize, they do show for a few practical, real-world cases that a swarm relaxation strategy can shorten the time to obtain results by a factor of several hundred, in return for an increased computational cost of about a factor of 3.  

\section{Discussion}

In addition to a dramatic decrease in the time required to obtain results, another significant advantage of the swarm relaxation approach is the quality of the results, including their error estimates, and the ease with which they are evaluated.  
All the microstates that contribute to the final results in a swarm approach are, by construction, completely independent.  
The quality of the estimates for equilibrium properties is thus very high, since they are formed as pure ensemble averages. 
{\color{black} While we have focussed here on bulk average properties such as $\rho$ and $U$, all observables available from a SLR can be readily computed from a swarm ensemble, including structural measures such as radial distribution functions, and quantities such as the specific heat that are based on fluctuations occurring within the ensemble.}
Also, since there is no need to estimate the time separation between independent microstates, as in a SLR, the evaluation of statistical error is straightforward and robust.  A swarm approach is therefore a good choice for studies requiring high-precision results, with rigorously defined error.

We emphasize that the swarm relaxation strategy does not resolve the fundamental physical challenges associated with the equilibration of complex systems.  Users of the present approach must still be watchful for the effects of metastable states, and of slowly relaxing collective degrees of freedom.  The approach does provide opportunities for checking for these effects, for example, testing for the presence of distinct metastable states by looking for divergent behavior in subsets of the swarm trajectories.  If the presence of a slow degree of freedom is suspected, it would be best to check swarm results against a test case using a SLR, especially if the system under study is new.

Regarding the definition of the autocorrelation functions used here, there are of course other choices that may serve just as well, or even better, for assessing the relaxation of the system to equilibrium.  
In particular, the time decay of the intermediate scattering function has long been used as a benchmark for quantifying relaxation in bulk liquids and glasses~\cite{berthier}.  When available, such additional measures of decorrelation can be used in a swarm approach to check for subtle, slowly relaxing degrees of freedom.
Here, we have focussed on the autocorrelation functions obtained from the time series for the same observables (e.g. $\rho$ and $U$) used to compare the swarm results to a SLR.  We do so for simplicity, and to show that when $M$ is large enough, any observable can be used to monitor the time evolution of $\tau$ as the system approaches equilibrium.

We have shown that $M=10^3$ is sufficient to make our strategy both efficient and straightforward to implement.  Smaller values of $M$ may also be used, at the cost of decreased precision in the estimates for equilibrium properties and for characteristic time scales such as $\tau$.  In particular, if the behavior of $\tau$ as a function of $t$ [see Fig.~\ref{tstop}(a)] is too noisy, then reliable estimates for $t_{\rm stop}$ 
become difficult to obtain.  Tests using sub-ensembles of our swarm runs suggest that $M=250$ is an approximate lower bound for obtaining accurate equilibrium properties, while simultaneously ensuring that a useful estimate for $t_{\rm stop}$ can be made from the behavior of $\tau(t)$.

{\color{black} In situations where $M\sim 10^3$ computing processors are not available for concurrent use, the swarm strategy can still be implemented, since the individual runs are independent and can run asynchronously.  Furthermore, the computational workload in a swarm approach takes the form of a large number of short runs.  Our experience when running asynchronously on a shared facility is that excellent throughput is achieved, since the runs fill usage gaps between larger and longer computing jobs.  

We also note that the swarm approach can be modified by extending each run so as to produce a sequence of independent configurations, appropriately separated in time.  In this case, observables are evaluated from a combination of ensemble and time averaging.  The balance between the two kinds of averaging can be tuned to best fit the available computing resources, bearing in mind that such a hybrid approach does not minimize the wall-clock time, and complicates the error analysis, relative to a pure swarm strategy.}

Finally, we point out the conceptual connections between our work and studies of physical aging in glassy systems.  The swarm procedure used here is the same as that commonly used in simulations to study the aging of material properties in a glass subjected (e.g.) to a jump in $T$.  The only difference is that here the destination equilibrium state can be reached, and that the characteristic time scales are much shorter than those normally studied in aging.  In particular, we draw the reader's attention to Dyre's recent analysis of the Narayanaswamy theory for physical aging, in which the ``material time" is unambiguously related to the system's mean-square-displacement in configuration space~\cite{dyre}.
The variation of $\tau$ with $t$ shown in Fig.~\ref{tstop} is a proxy measure of the material time in our test systems as they approach equilibrium.  It would be interesting for future work to assess the swarm relaxation strategy within the framework of Dyre's analysis. 

To summarize, the practicality of the swarm relaxation strategy rests on two observations:  (i)  The time required to generate independent microstates during a single long run is comparable to the time required to bring a single short run into equilibrium.  (ii) When the swarm is large enough, the attainment of equilibrium can be confirmed within a time that is not much longer than the equilibration process itself.  So long as these two observations hold, the present strategy is an effective way to ``trade processors for time".  
When computational facilities having $10^3$ or more processors are available, and when time is of the essense, the swarm relaxation strategy is an effective way to rapidly generate high-quality results with robustly defined statistical error.

\begin{acknowledgements}
RKB, ISV and PHP thank NSERC for support.  
Computational resources were provided by ACEnet.  
\end{acknowledgements}

{\color{black}
\bigskip
\centerline{\bf Author Contributions}
\medskip
All authors contributed to the development of this study and the interpretation of the results.  
FS provided the Monte Carlo code for the bulk liquid ST2 simulations, which were carried out by PHP.  SMAM carried out the TIP4P/2005 nanodroplet simulations.  
PHP drafted the manuscript, with input from all authors.
}

\appendix*
\section{Fluctuations of the autocorrelation function}

Here we show that the fluctuations of the autocorrelation function $C_x(t_0,t)$ have a standard deviation $\sigma_C=M^{-1/2}$, when $C_x$ approaches zero.

Let $X(t)$ represent the discrete set of $M$ random variables $\{x(i,t)\}$ for various $i$ at fixed $t$. 
Similarly, let $\delta X(t)$ represent the discrete set of $M$ random variables $\{x(i,t)-\langle x(t)\rangle \}$.  
The variance of $X(t)$ can be written in a number of ways:
\begin{eqnarray}
{\rm Var}[X(t)]&=&\sigma^2(t) \nonumber \\
&=&{\bigg\langle \Big[x(i,t)-\big\langle x(t)\big\rangle\Big]^2\bigg\rangle} \nonumber \\
&=& \Big\langle\big[\delta X(t)\big]^2 \Big\rangle  \nonumber \\
&=& \Big\langle\big[X(t)\big]^2 \Big\rangle  -   \big\langle X(t)\big\rangle^2.
\label{variance}
\end{eqnarray}

In this notation,
\begin{equation}
C_x(t_0,t) = \frac{\big\langle \delta X(t_0)\,\, \delta X(t) \big\rangle}{\sigma(t_0)\,\sigma(t)}.
\end{equation}
The fluctuations of $C_x$ are quantified by ${\rm Var}[C_x(t_0,t)]=\sigma^2_C$.  Using standard identities for the variance, we have,
\begin{eqnarray}
{\rm Var} [C_x(t_0,t)] &=& {\rm Var} \Bigg[ \frac{\big\langle \delta X(t_0)\,\, \delta X(t) \big\rangle}{\sigma(t_0)\,\sigma(t)}\Bigg] \nonumber \\
                              &=& \frac{{\rm Var}\Big[\big\langle \delta X(t_0)\,\, \delta X(t) \big\rangle\Big]}{\sigma^2(t_0)\,\sigma^2(t)} \nonumber \\
                              &=& \frac{{\rm Var}\big[ \delta X(t_0)\,\, \delta X(t) \big]}{M\, \sigma^2(t_0)\,\sigma^2(t)}.
\label{A3}                              
\end{eqnarray}
Using the last equality of Eq.~\ref{variance} we can write,
\begin{eqnarray}
{\rm Var}\big[ \delta X(t_0)\,\, \delta X(t) \big] &=& \Big\langle\big[\delta X(t_0)\big]^2 \big[\delta X(t)\big]^2 \Big\rangle     \nonumber \\  
&&-   \big\langle\delta X(t_0)\,\, \delta X(t)\big\rangle^2
\label{A4}
\end{eqnarray}
For sufficiently large $\Delta t$, $\delta X(t_0)$ and $\delta X(t)$ become independent, and $C_x\to 0$.  In this case, the first term on the right-hand side of Eq.~\ref{A4} reduces to,
\begin{eqnarray}
\Big\langle\big[\delta X(t_0)\big]^2 \big[\delta X(t)\big]^2 \Big\rangle &=& \Big\langle\big[\delta X(t_0)\big]^2 \Big\rangle \Big\langle\big[\delta X(t)\big]^2 \Big\rangle \nonumber \\
&=& \sigma^2(t_0)\,\sigma^2(t),
\end{eqnarray}
and the second term vanishes,
\begin{eqnarray}
\big\langle\delta X(t_0)\,\, \delta X(t)\big\rangle^2 = \big\langle\delta X(t_0) \big\rangle^2 \big\langle\delta X(t)\big\rangle^2 =0, 
\label{A6}
\end{eqnarray}
because by definition $\langle\delta X(t_0)\rangle=\langle\delta X(t)\rangle=0$.  Combining Eqs.~\ref{A3}-\ref{A6}, we obtain,
\begin{eqnarray}
{\rm Var} [C_x(t_0,t)] = M^{-1}.
\end{eqnarray}    
Therefore, the standard deviation $\sigma_C$ of fluctuations of $C_x$ as $C_x\to 0$ is,
\begin{eqnarray}
\sigma_C = M^{-1/2}.
\end{eqnarray}

\end{document}